\documentclass[aps, twocolumn, notitlepage, superscriptaddress, prb]{revtex4-1}

\usepackage[english]{babel}
\usepackage{mathtools}
\usepackage{graphicx}
\usepackage{epsfig}
\usepackage{wasysym}
\usepackage{amsfonts}
\usepackage{bm}
\usepackage{color}
\usepackage[normalem]{ulem}
\usepackage{cancel}
\usepackage{enumitem}

\usepackage{microtype}
\usepackage{braket}
\usepackage{multirow}

\DeclareMathOperator{\Tr}{\mbox{Tr}}

\newcommand{\pdftitle}{Ware_Abanin_Vasseur_2020}

\newcommand{\Dthree}{\ensuremath{\text{D}_3}\,}
\newcommand{\Dk}[1]{\ensuremath{\text{D}_{#1}}}
\newcommand{\Otwo}{\ensuremath{\text{O}(2)}\,}
\newcommand{\sutwo}{\ensuremath{\text{SU}(2)}\,}
\newcommand{\sutwok}[1]{\ensuremath{\text{SU}(2)_{#1}}}
\newcommand{\sothreek}[1]{\ensuremath{\text{SO}(3)_{#1}}}
\newcommand{\Fib}{\ensuremath{\text{Fib}}}
\newcommand{\Ising}{\ensuremath{\text{Ising}}}

\newcommand{\iA}{\ensuremath{\mathcal A}}

\usepackage{amsmath}
\usepackage{amsfonts}
\usepackage{amssymb}
\usepackage{braket}

\usepackage[hidelinks]{hyperref}
\hypersetup{
    pdftitle={\pdftitle},
    pdfauthor={Brayden Ware},
    colorlinks=true,
    linkcolor=black,
    citecolor=blue,
    filecolor=magenta,
    urlcolor=blue
}

\newcommand{\x}{\mathbf{x}}

\def\bra#1{\mathinner{\langle{#1}|}}
\def\ket#1{\mathinner{|{#1}\rangle}}
\def\braket#1{\mathinner{\langle{#1}\rangle}}

\usepackage{calc}
\newlength{\depthofsumsign}
\newcommand{\nsum}[1][1.4]{
\mathop{%
    \raisebox
        {-#1\depthofsumsign+1\depthofsumsign}
        {\scalebox
            {#1}
            {$\displaystyle\sum$}%
        }
}
}

\usepackage{tikz}

\usetikzlibrary{trees,arrows,positioning,arrows.meta}
\usetikzlibrary{decorations.pathreplacing,decorations.markings}
\usetikzlibrary{calc}
\tikzset{
  on each segment/.style={
    decorate,
    decoration={
      show path construction,
      moveto code={},
      lineto code={
        \path [#1]
        (\tikzinputsegmentfirst) -- (\tikzinputsegmentlast);
      },
      curveto code={
        \path [#1] (\tikzinputsegmentfirst)
        .. controls
        (\tikzinputsegmentsupporta) and (\tikzinputsegmentsupportb)
        ..
        (\tikzinputsegmentlast);
      },
      closepath code={
        \path [#1]
        (\tikzinputsegmentfirst) -- (\tikzinputsegmentlast);
      },
    },
  },
  mid arrow reverse/.style={postaction={decorate,decoration={
        markings,
        mark=at position .6 with {\arrow[#1]{stealth}}
      }}},
  mid arrow/.style={postaction={decorate,decoration={
        markings,
        mark=at position .4 with {\arrowreversed[#1]{stealth}}
      }}},
  early arrow/.style={postaction={decorate,decoration={
        markings,
        mark=at position .25 with {\arrowreversed[#1]{stealth}}
      }}},
  later arrow/.style={postaction={decorate,decoration={
        markings,
        mark=at position .45 with {\arrowreversed[#1]{stealth}}
      }}},
}
\tikzstyle arrowstyle=[scale=2,black]
 
\begin{document}

\title{Perturbative instability of non-ergodic phases in non-Abelian quantum chains}

\author{Brayden Ware}
\affiliation{Department of Physics, University of Massachusetts, Amherst, MA 01003, USA}
\author{Dmitry Abanin}
\affiliation{Department of Theoretical Physics, University of Geneva, Geneva, Switzerland} 
\author{Romain Vasseur}
\affiliation{Department of Physics, University of Massachusetts, Amherst, MA 01003, USA}
\begin{abstract}

An important challenge in the field of many-body quantum dynamics is to identify non-ergodic states of matter beyond many-body localization (MBL). Strongly disordered spin chains with non-Abelian symmetry and chains of non-Abelian anyons are natural candidates, as they are incompatible with standard MBL. In such chains, real space renormalization group methods predict a partially localized, non-ergodic regime known as a \emph{quantum critical glass} (a critical variant of MBL). This regime features a tree-like hierarchy of integrals of motion and symmetric eigenstates with entanglement entropy that scales as a logarithmically enhanced area law. We argue that such tentative non-ergodic states are perturbatively unstable using an analytic computation of the scaling of off-diagonal matrix elements and accessible level spacing of local perturbations. Our results indicate that strongly disordered chains with non-Abelian symmetry display either  spontaneous symmetry breaking or ergodic thermal behavior at long times. We identify the relevant length and time scales for thermalization: even if such chains eventually thermalize, they can exhibit non-ergodic dynamics up to parametrically long time scales with a non-analytic dependence on disorder strength.

\end{abstract}
 
\maketitle
\section{Introduction}
\label{sec:intro}

The investigation of isolated quantum systems and whether they ``self-thermalize" has been the focus of much theoretical and experimental work in recent years~\cite{schreiberObservationManybodyLocalization2015,smithManybodyLocalizationQuantum2016,choiExploringManybodyLocalization2016,Lukin256,2019arXiv191006024C}. In particular, certain one dimensional systems with strong quenched randomness have been shown to completely evade thermalization. These \emph{many-body localized} (MBL) phases protect quantum coherence and remain out of equilibrium at any (effective) temperature and for infinite times~\cite{baskoMetalInsulatorTransition2006,doi:10.1146/annurev-conmatphys-031214-014726,Vasseur_2016,RevModPhys.91.021001}. 
Theoretically, the hallmark of MBL is the emergence of a complete set of exact local integrals of motion (LIOMs)~\cite{serbynLocalConservationLaws2013,husePhenomenologyFullyManybodylocalized2014,rosIntegralsMotionManybody2015,chandranConstructingLocalIntegrals2015a,imbrieManyBodyLocalizationQuantum2016}. The existence of LIOMs has been used to establish that the MBL behavior is not merely a non-equilibrium regime but a fully stable eigenstate phase of matter. Further, the characteristic properties of the MBL phase can be described as a consequence of the LIOMs; in particular, the area-law scaling of entanglement typical of gapped ground states applies to eigenstates throughout the spectrum~\cite{bauerAreaLawsManybody2013, serbynLocalConservationLaws2013}. 
Additionally, these eigenstates can exhibit quantum orders usually restricted to zero-temperature, including symmetry-breaking, topological and symmetry protected topological (SPT) order~\cite{huseLocalizationprotectedQuantumOrder2013,bauerAreaLawsManybody2013, kjallManyBodyLocalizationDisordered2014, Bahri:2015aa, chandranManybodyLocalizationSymmetryprotected2014}.

Previous work has shown that the existence of a complete set of LIOMs is incompatible with protected degenerate excitations such as those that appear in non-Abelian symmetric phases, in topological orders with non-Abelian anyons, or in symmetry enriched topological order with excitations that carry projective representations of the symmetry~\cite{potterSymmetryConstraintsManybody2016,chandranManybodyLocalizationSymmetryprotected2014}.
Thus, these systems cannot exhibit MBL. This motivates the question of whether there are stable non-ergodic eigenstate phases outside the MBL paradigm, i.e. with some degree of localization but without a full set of LIOMs. 
If not, the fate of these systems is either thermalization or full localization but with eigenstates that spontaneously break the symmetry to an Abelian subgroup~\cite{potterSymmetryConstraintsManybody2016,vasseurParticleholeSymmetryManybody2016a,prakashEigenstatePhasesFinite2017, friedmanLocalizationprotectedOrderSpin2018}. 

One promising proposal for such non-ergodic eigenstate phases are a class of states known as \emph{quantum critical glasses} (QCGs)~\cite{vasseurQuantumCriticalityHot2015,voskManyBodyLocalizationOne2013,kangUniversalCrossoverGroundstate2017,parameswaranEigenstatePhaseTransitions2017}.
These phases have eigenstates which are as localized as possible while preserving the non-Abelian symmetry~\cite{protopopovEffectSUSymmetry2017}. They feature a hierarchical set of integrals of motion at all length scales, most of which are local, while a few involve a finite fraction of spins in the system.
Like the LIOMs for MBL, the existence of the hierarchical IOMs strongly constrains the dynamical properties of QCGs. For example, the entanglement growth after a quench in the presence of these IOMs scales as $\sim \log^{1/\psi} t$ with $\psi<1$, as compared to the scaling $\sim \log t$ for MBL phases~\cite{znidaricManybodyLocalizationHeisenberg2008, bardarsonUnboundedGrowthEntanglement2012,serbynUniversalSlowGrowth2013}.  
Similarly, eigenstates have logarithmically scaling entanglement~\cite{refaelCriticalityEntanglementRandom2009} instead of area law entanglement for MBL. 

An approximate construction for the QCG states is provided by the strong disorder renormalization group for excited states (RSRG-X)~\cite{voskManyBodyLocalizationOne2013, voskDynamicalQuantumPhase2014, pekkerHilbertGlassTransitionNew2014, vasseurQuantumCriticalityHot2015,agarwalEnsuremathAlphaNoise2015,Monthus_2016,PhysRevB.93.104205, PhysRevB.94.014205}, building on ground-state RSRG methods~\cite{PhysRevLett.69.534, PhysRevB.50.3799,PhysRevB.51.6411}. This construction yields at lowest order a picture of QCG states as tree tensor networks with irregular, disorder realization dependent shapes with an IOM associated with each node of the tree.
While approximate, this construction is increasingly accurate as disorder strength is increased, as can be confirmed on (small) finite size systems with exact diagonalization~\cite{kangUniversalCrossoverGroundstate2017}. Thus, a natural starting place for establishing the existence of QCG phases is to determine if RSRG-X reliably approximates the true eigenstates of strongly disordered spin chains in the thermodynamic limit. If it does not, understanding the microscopic processes that cause the failure of RSRG-X will provide insight into whether these systems are non-ergodic or if they thermalize.



Previous efforts have shown that the LIOMs of MBL are stable at sufficiently strong disorder, using both perturbative analyses and non-perturbative considerations such as the inclusion of thermal regions~\cite{imbrieManyBodyLocalizationQuantum2016,roeckManybodyLocalizationStability2017,baskoMetalInsulatorTransition2006,serbynCriterionManyBodyLocalizationDelocalization2015}. 
At weaker disorder, the presence of many collective many-body resonances destabilizes the LIOMs and melts the MBL phase into a incoherent thermal liquid~\cite{voskTheoryManyBodyLocalization2015,potterUniversalPropertiesManyBody2015,serbynCriterionManyBodyLocalizationDelocalization2015, PhysRevX.7.021013, PhysRevLett.119.110604, PhysRevLett.121.140601, PhysRevLett.122.040601, PhysRevB.99.094205, PhysRevB.99.224205,2020arXiv200604825M}. The non-perturbative stability of MBL remains an important question in various contexts~\cite{deroeckStabilityInstabilityDelocalization2017,PhysRevLett.119.150602,PhysRevB.99.205149,PhysRevB.99.134305,PhysRevResearch.2.033262}, including in dimension $d>1$ and with long-range interactions.

The success or failure of RSRG-X for QCG phases can also be understood through the lens of resonances. When the distribution of couplings in the RSRG-X process flows to an increasingly broad distribution, local resonances caused by collisions of neighboring couplings are increasingly unlikely. This scenario occurs in  the \emph{infinite randomness fixed points} discussed in Refs.~\onlinecite{vasseurQuantumCriticalityHot2015,voskManyBodyLocalizationOne2013}.
An extension of this analysis to collisions of more distant couplings was considered in Ref.~\onlinecite{voskManyBodyLocalizationOne2013}, which found that these resonances were also irrelevant near the infinite randomness fixed point.
Other studies found scenarios where local resonances led to the breakdown of RSRG-X, as in Ref.~\onlinecite{vasseurQuantumCriticalityHot2015} which considered \sutwok{k} anyon chains in the limit $k \to \infty$, considered as a proxy for \sutwo symmetric chains, and in Ref.~\onlinecite{vasseurParticleholeSymmetryManybody2016} which considered \Otwo symmetric spin chains (a.k.a fermionic chains with particle-hole symmetry $\Otwo = {\rm U}(1) \rtimes {\mathbb Z}_2$, where the non-Abelian semi-direct product structure reflects the nontrivial action of a ${\mathbb Z}_2$ particle-hole symmetry on the conserved $\rm{U}(1)$ charge).

The resonances considered in these analyses all involve processes that couple a few IOMs. To probe stability against many-body resonances, Refs.~\onlinecite{protopopovEffectSUSymmetry2017,protopopovNonAbelianSymmetriesDisorder2019} introduced a technique to analyze multi-spin processes that mix states that differ in many IOMs~\cite{PhysRevB.92.104202}. 
These resonances involve processes that couple approximate IOMs produced by RSRG-X at all levels in the hierarchy, as indicated in Fig.~\ref{fig:treecartoon}, and thus are \emph{global} resonances involving a large fraction of the spins in the spin chain. 
Applying their method to \sutwo symmetric spin chains, they discovered a proliferation of these collective many-body resonances in the thermodynamic limit, even with arbitrarily strong disorder.
Unlike in the MBL phase, where resonances are likely to be spatially separated and involve disjoint sets of LIOMs, the resonances they identified involve overlapping sets of IOMs, which likely ``percolate'' and drive the system to a thermal (ergodic) phase.

The studies above are all consistent with a general picture where RSRG-X breaks down for chains with continuous non-Abelian symmetries but is successful for strongly disordered spin chains with discrete non-Abelian symmetry or for non-Abelian anyon chains. 
However, no existing study of the latter has considered multispin resonant processes as Refs.~\onlinecite{protopopovEffectSUSymmetry2017,protopopovNonAbelianSymmetriesDisorder2019} did for \sutwo symmetric chains.
Numerical studies~\cite{friedmanLocalizationprotectedOrderSpin2018,kangUniversalCrossoverGroundstate2017} using exact diagonalization have been done for some examples of such systems, but generally speaking the effects of rare collective resonances are not expected to show up on the length or time scales accessible to these computations~\cite{191104501Distinguishing,191107882Can}. A quantitative understanding of the resonances would give insight into the scales needed to study thermalization either numerically or experimentally.

In this paper, we carry out an analysis of multispin resonant processes for discrete non-Abelian chains and find that resonances driven by these processes proliferate in large enough systems. To show that the processes we identify indeed cause many resonances, and to estimate the associated length scales on which thermalization occurs, we use a combination of explicit computations and analytic arguments. Our argument is constructed as follows: In Sec.~\ref{sec:criteria} and Sec.~\ref{sec:resonant_processes}, we identify the resonant processes and describe the criteria we use for determining the stability. In Sec.~\ref{sec:matrixelementformula}, we set up formalism for the quantitative analysis of resonant multispin processes. We express relevant matrix elements of local operators between RSRG-X states in terms of Clebsch-Gordan tensors, deriving a compact analytic formula. In Sec.~\ref{sec:numericmatrixelements}, we compute numerically the number of non-vanishing matrix elements and their distribution for the resonant processes. In Sec.~\ref{sec:randommatrixproducts} we show that these matrix element computations can be mapped to a transfer matrix-like calculation, so that the size of the matrix elements of local operators between typical QCG states is captured by the scaling of a random product of transfer matrices. This allows us to extract the asymptotic scaling of the matrix elements by estimating the Lyapunov exponent controlling the growth of the random matrix product. We carry out this computation for a number of discrete non-Abelian groups and anyon theories, finding in each case a number of resonances scaling as a power-law in system size, and that the exponents match explicit counting of resonances. In Sec.~\ref{sec:constraintsonexponents} we argue that the same result occurs generally for any non-Abelian group and for all anyon chains with exceptions for Majorana and parafermion chains.
Finally, in Sec.~\ref{sec:directcomputation} we explicitly count the number of resonances produced when locally perturbing strongly disordered Fibonacci chains (a simple example of anyonic chain). This count of resonances matches the computations of Sec.~\ref{sec:sizeandnumber} and confirms the scenario described in Sec.~\ref{sec:resonant_processes}.

\section{Criteria for perturbative instability of QCG}
\label{sec:criteria}

An unusual feature of the MBL phase is that the eigenstates deep within the many-body spectrum, where the level spacing is exponentially small in the system size, are stable to local perturbations. This arises because states connected by a sizeable matrix element of a local perturbation differ in the value of a small number of LIOMs, and thus differ by a constant-sized gap except in rare cases. 
Nearby states in the spectrum can also be mixed by local perturbations, but because they differ in many LIOMs, this mixing occurs at high orders in perturbation theory and thus with matrix elements that are thus exponentially suppressed in the number of flipped LIOMs. 
The many-body Thouless parameter~\cite{serbynCriterionManyBodyLocalizationDelocalization2015}
\begin{equation}
\label{eq:thouless}
\mathcal{G}^V_{ab} = \log \left| \frac{V_{ab}}{E_a - E_b} \right |
\end{equation}
captures the ability of a local perturbation $V$ to mix eigenstates $\ket{a}, \ket{b}$ with energies $E_a, E_b$.
As shown in Ref.~\onlinecite{serbynCriterionManyBodyLocalizationDelocalization2015}, the MBL phase is characterized by $\mathcal{G}_{n, n+1} \propto -L$ for typical neighboring eigenstates $\ket{n}, \ket{n+1}$ in a spin chain of size $L$, while the ergodic phase features $\mathcal{G}_{n, n+1} \propto +L$.

We can use these criteria to characterize the perturbative stability of tree eigenstates (say, produced by RSRG-X) in strongly disordered non-Abelian chains against many-body collective resonances. Let $\{\ket{a}\}$ represent a basis of approximate RSRG-X eigenstates of such a chain with a Hamiltonian $H$, and let 
$$\text{Diag}(H) = \sum_a E_a \ket{a} \bra{a}$$ be the diagonal part of $H$ in the basis of such approximate eigenstates.
We argue that certain local operators $V$ exhibit
\begin{equation}
\mathcal{G}^V_{n, n+1} \sim \gamma' \log \frac{L}{L_0},
\label{eq:gammaprime}
\end{equation}
with a positive constant $\gamma'$.
To connect that to the accounting of resonances, note that 
each occurence of $
\log \lambda + \mathcal{G}^V_{ab} > 0
$
is a resonance in 
$$H' = \text{Diag}(H) + \lambda V;$$
and thus the perturbation $\lambda V$ effectively hybridizes nearby tree eigenstates whenever $\lambda > \lambda_c = (L/L_0)^{-\gamma'}$.
Whenever $\gamma'>0$, the size of perturbation that destabilizes the tree eigenstates goes to $0$ in the thermodynamic limit. We use this as our criteria for perturbative instability.

This perturbative instability naturally leads to a breakdown in RSRG-X above disorder strength-dependent length and time scales. 
At an RSRG-X step decimating a bond coupling of strength $J_i$, terms in the Hamiltonian are discarded which are local operators of strength $\delta J$ set by the neighboring couplings $\delta J \sim J_{i \pm 1}$. 
At strong disorder, $\delta J / J \sim 1/W$ with $W$ the disorder strength.
The cumulative effect of these RSRG-X errors can be interpreted in our perturbative framework by setting $\lambda V = H - \text{Diag}(H)$.
By the above analysis, we can thus extract a length scale 
$$
L_{\text{th}} \sim L_0 \left(\frac{J}{\delta J} \right)^{1/\gamma'},
$$ above which many-body resonances proliferate and lead to thermalization.
A more detailed analysis of the thermalization length scale was carried out in Ref.~\onlinecite{protopopovNonAbelianSymmetriesDisorder2019} for the \sutwo symmetric disordered Heisenberg chain, yielding a result consistent with this picture.

This thermalization length scale $L_{\text{th}}$ can be converted to a thermalization time scale $t_{\text{th}}$ using the dynamical scaling~\cite{vasseurQuantumCriticalityHot2015} of QCG $\log t \sim L^{\psi}$ for some universal exponent $\psi<1$, valid for $t \lesssim t_{\text{th}}$. This yields a stretched-exponentially long time scale
\begin{equation} \label{eqtth}
 t_{\text{th}} \sim t_0 \ {\rm exp}\left({C \left(\frac{J}{\delta J} \right)^{\psi/\gamma'}} \right).
 \end{equation}
 For $t \ll  t_{\text{th}} $, the quantum dynamics is non-ergodic and well captured by RSRG-X, while at long times $t \ll  t_{\text{th}}$, many-body resonances proliferate and the system thermalizes.  
 Note that even though the instability to thermalization through many-body resonances is {\rm perturbative}, this thermalization timescale has a non-analytic dependence on $\delta J$. In particular, this time scale can be extremely long at strong enough disorder, making the non-ergodic behavior of QCG very robust even though thermalization eventually takes over. 
 
 In the following, we will show that Eq.~\eqref{eq:gammaprime} holds, and compute the exponent $\gamma' >0$.
 
\section{Structure of the resonant processes}
\label{sec:resonant_processes}
In this Section we identify the resonant processes responsible for the perturbative instability of QCGs indicated by Eq.~\ref{eq:gammaprime} in strongly disordered non-Abelian spin chains.
The critical feature of such non-ergodic QCG states are the non-local IOMs whose presence is forced by the symmetry.
As we will show, local perturbations can couple these non-local IOMs to a large number of other IOMs. To understand this, we need to briefly review the mechanics of RSRG-X.

At each stage of the RSRG-X procedure, a pair of neighboring spins (or anyons) coupled by the strongest bond in the system is chosen. As the interaction between these spins is typically much larger than to their neighbors, the spectrum generically splits into sectors in which the two spins --- which transform under the non-Abelian symmetry as representations $\mathbf{a}$ and $\mathbf{b}$ --- transform together as a single irreducible representation $\mathbf{c}$ taken from the decomposition $\mathbf{a} \times \mathbf{b} = \sum_c N_{ab}^c \mathbf{c}$.
If the dimension $d_{\mathbf{c}}$ of the irrep $\mathbf{c}$ is bigger than $1$, the two spins can be replaced by a single effective ``spin'' transforming as $\mathbf{c}$ with renormalized coupling to its neighbors. If $d_{\mathbf{c}}=1$, $\mathbf{a}$ and $\mathbf{b}$ form a singlet and decouple completely, and there is an effective coupling between their neighbors that can be computed within 2nd order perturbation theory.
If $\mathbf{a}, \mathbf{b}$ are irreducible representations, the corresponding lowest order wavefunctions of the spins are thus completely constrained to be the Clebsch-Gordan tensors for the fusion product $\mathbf{a} \times \mathbf{b} \to \mathbf{c}$. If the initial spins do not transform as irreducible representations of the non-Abelian symmetry, the RSRG-X process replaces pairs of these spins with effective spins consisting of an irreducible multiplet of states --- and thus Clebsch-Gordan tensors result for subsequent decimations at larger length scales. Some useful properties of Clebsch-Gordan tensors are reviewed in Appendix~\ref{sec:treestates}.

By repeating the RSRG-X decimation process until no spins are left, a complete orthogonal basis of states indexed by the outcomes of each fusion is generated. Each state is represented at lowest order by a tree tensor network of Clebsch-Gordan tensors with an irregular, disorder-dependent and energy-dependent tree shape. 
These tree-shaped eigenstates can be thought of as the least-entangled possible states compatible with the non-Abelian structure~\cite{protopopovEffectSUSymmetry2017}.

The energy dependence in the tree shape occurs because the choice of fusion outcome at one step can change the order of decimations at later steps --- but only if the associated energy scales of those later steps were close enough. In the limit of strong disorder, this happens less and less frequently. 
Thus, for the purposes of evaluating our perturbative instability criteria, we can focus entirely on states with the same tree shape (for each given disorder instance). A quantitative analysis on how often RSRG-X histories diverge in tree shape was made in Ref.~\onlinecite{kangUniversalCrossoverGroundstate2017} with the same conclusion. 

\begin{figure}
    \centering
    \includegraphics[width=0.9\linewidth]{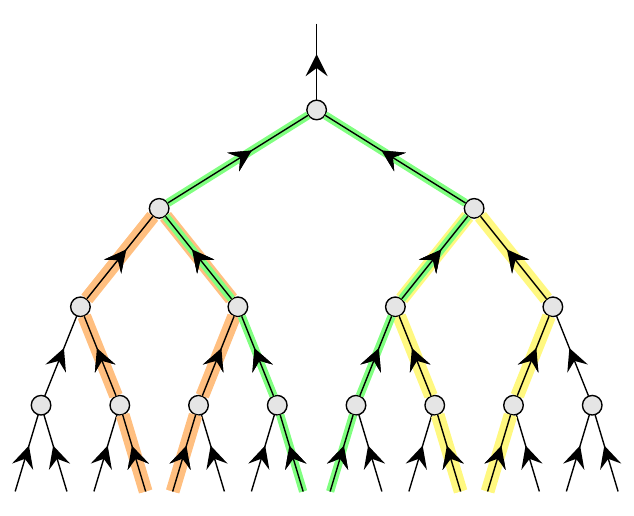}
    \caption{Symmetric operators acting on two neighboring sites can only mix states whose IOMs differ only along the direct connecting path (highlighted).}
    \label{fig:treecartoon}
\end{figure}

We will thus do our analysis of the resonances of $\text{Diag}(H) + \lambda V$
using the complete basis of fusion states with a fixed tree shape. 
With this simplification, each edge of the tree corresponds to an approximately conserved IOM operator which projects to a fixed irrep at that edge. As illustrated in Fig.~\ref{fig:treecartoon}, local symmetric operators applied to a tree state give a superposition of tree states that differ only their irrep labels in part of the tree --- particularly, the edges of the tree above the microscopic spins on which the operator acts but below the point where those spins fuse into a single irrep. The number $D$ of IOMs that are effected for a given local operator scales with the depth of the tree in the region of that operator.
Our key argument for showing that resonances proliferate developed below is based on understanding how the matrix elements change as a function of $D$. In the computation of the following sections, we show that the number of non-zero matrix elements of a local operator scales exponentially in $D$, and the size of such matrix elements decays exponentially in $D$. As bonds corresponding to the deepest cuts through the tree have a tree-depth scaling logarithmically in the system size, these scalings translate to power-law scaling in system size.
The basic mechanism for the proliferation of resonances is that the number of connected states scales faster with $D$ than the size of the matrix elements decays, leading to nearby level spacings between such states that are smaller than the matrix elements~\cite{protopopovEffectSUSymmetry2017}.

In what follows, we will use explicit computation of matrix elements to support these statements. 

\section{Exact formula for matrix elements between tree states}
\label{sec:matrixelementformula}

A benefit of our simplification to a fixed tree shape is that we can write an explicit formula for the matrix elements of local symmetric operators that allow us to understand the scaling of their size analytically rather than through brute-force numerics.
To derive this formula, we use a special basis of local operators that act on two adjacent spins transforming as irreps ${\bf r_1, r_2}$ which are formed from two Clebsch-Gordan tensors:
\begin{center}
\begin{tikzpicture}[on grid,
        edge/.style={draw,postaction={on each segment={early arrow=arrowstyle}}},
        lateredge/.style={draw,postaction={on each segment={later arrow=arrowstyle}}}]
    \def \l{1.5}
    \def \pdl{1.}
    \def \pds{0.5}
    \def \nd{0.5}
    \def \td{0.1}
    \node (center) {};
    \node[left = \l of center] {$O^q_{r_1, r_2} =$};
    \node[above left = 0.5*\nd and \nd of center] (a) [circle, draw] {};
    \node[above = \pds of a] (cu) {};
    \node[below = \pdl of a] (cd) {};
    \path[edge] (cu.north)--(a) node [midway, left=\td] {${\bf r_1}$};
    \path[lateredge] (a) -- (cd.south) node [midway, left=\td] {${\bf r_1}$};
    \node[below right = 0.5*\nd and \nd of center, circle, draw] (b) {};
    \node[above = \pdl of b] (du) {};
    \node[below = \pds of b] (dd) {};
    \path[lateredge] (du.north) -- (b) node [midway, right=\td] {${\bf r_2}$};
    \path[edge] (b)--(dd.south) node [midway, right=\td] {${\bf r_2}$};
    \path[red, edge] (a) -- node [midway, above=\td] {${\bf q}$} (b);
    \end{tikzpicture}
\end{center}
Here, ${\bf q}$ labels an irrep such that $N_{q r_1}^{r_1}, N_{q r_2}^{r_2} >0$. 
All symmetric operators acting on two spins can be written as a linear combination of these operators, so the matrix element for a generic two-site operator comes for free by decomposing it in this basis.
Operators on two anyons ${\bf r_1}, {\bf r_2}$ can also be parameterized in the same way, now with ${\bf q}$ labeling an anyon label --- but in that case there is no interpretation as a contraction of two Clebsch-Gordan tensors. See Appendix~\ref{sec:treestates} for more details on anyonic Hilbert spaces.
As an example, consider an \sutwo spin chain built from spins which tranform as the irreps  ${\bf r_1}, {\bf r_2} = {\bf \frac12}$. The two allowed values of ${\bf q}$ are the spin-${\bf 0}$ and spin-${\bf 1}$ irrep, which correspond to the identity operator and the $\vec{S} \cdot \vec{S}$ Heisenberg coupling, respectively.
The `spin' ${\bf q}$ transferred between the sites plays an important role in the formula below. 

To compute the action of such an operator on tree states, let us label the edges in the relevant part of the tree geometry as follows:
\begin{center}
  \begin{tikzpicture}[level/.style={sibling distance=50mm/#1, level distance = 1cm},
                      level 1/.style={level distance = 1.cm},
                      level 2/.style={sibling distance=45mm},
                      level 5/.style={level distance = 1.cm},
                      level 6/.style={level distance = 1.cm},
                      edge from parent/.style={draw,postaction={on each segment={mid arrow=arrowstyle}}},
                      edge/.style={draw,postaction={on each segment={early arrow=arrowstyle}}},
                      lateredge/.style={draw,postaction={on each segment={later arrow=arrowstyle}}}]
    \def \ang{-30}
    \node (c) [rotate=0] {$\vdots$}
      child {node [circle,draw] (z) {}
        child {node [circle,draw] (a) {}
          child {node [rotate=\ang] (b) {$\vdots$}
          }
          child {node [circle,draw] (g) {}
            child {node [rotate=\ang] (s) {$\vdots$}
            }
            child {node [circle,draw] (i) {}
              child {node [rotate=\ang] (f) {$\vdots$}
              }
              child {node (h) {\iffalse h \fi }
              }
            }
          }
        }
        child {node [circle,draw] (j) {}
          child {node [circle,draw] (k) {}
            child {node [circle,draw] (m) {}
              child {node (d) {}
              }
              child {node [rotate=-\ang] (e) {$\vdots$}
              }
            }
            child {node [rotate=-\ang] (t) {$\vdots$}}
          }
        child {node [rotate=-\ang] (l) {$\vdots$}
        }
      }
    };
    \path (h) -- (i) node [midway, right=0.05cm] {${\bf r_1}$};
    \path (d) -- (m) node [midway, left=0.05cm] {${\bf r_2}$};
    \path (a) -- (z) node [midway, below=0.1cm] {${\bf b_m}$};
    \path (g) -- (a) node [midway, right=0.1cm] {${\bf b_2}$};
    \path (i) -- (g) node [midway, right=0.1cm] {${\bf b_1}$};
    \path (j) -- (z) node [midway, below=0.1cm] {${\bf d_n}$};
    \path (k) -- (j) node [midway, left=0.1cm] {${\bf d_2}$};
    \path (m) -- (k) node [midway, left=0.1cm] {${\bf d_1}$};
    \path (z) -- (c) node [midway, right=0.1cm] {${\bf c}$};
    \path (b) -- (a) node [midway, left=0.1cm] {${\bf a_3}$};
    \path (s) -- (g) node [midway, left=0.1cm] {${\bf a_2}$};
    \path (f) -- (i) node [midway, left=0.1cm] {${\bf a_1}$};
    \path (l) -- (j) node [midway, right=0.1cm] {${\bf e_3}$};
    \path (t) -- (k) node [midway, right=0.1cm] {${\bf e_2}$};
    \path (e) -- (m) node [midway, right=0.1cm] {${\bf e_1}$};

    \def \pdl{0.4}
    \def \pds{0.1}
    \def \nd{0.1}
    \def \ed{-0.1}
    \def \td{0.1}
    \node (center) at (0, -5.7){};
    \node[above left = \ed and \nd of center] (a) [circle, draw] {};
    \node[above = \pds of a] (cu) {};
    \node[below = \pdl of a] (cd) {};
    \path[edge] (cu.north)--(a) node [midway, left=\td] {}; 
    \path[lateredge] (a) -- (cd.south) node [midway, left=\td] {}; 
    \node[below right = \ed and \nd of center, circle, draw] (b) {};
    \node[above = \pdl of b] (du) {};
    \node[below = \pds of b] (dd) {};
    \path[lateredge] (du.north) -- (b) node [midway, right=\td] {}; 
    \path[edge] (b)--(dd.south) node [midway, right=\td] {}; 
    \path[red, edge] (a) -- node [midway, above=\td] {${\bf q}$} (b);
  \end{tikzpicture}
\end{center}
This shows only the relevant portion of the tree for an operator being applied to the spins labeled ${\bf r_1, r_2}$; other branches can exist below each of the ${\bf a_i, e_i}$ or otherwise only connected to this portion of the tree through ${\bf c}$. The number of branches on the left and right of the tree will be referred to as $m$ and $n$ respectively; the picture shows $m=n=3$.

The local operator only connects tree states with different values for the irreps corresponding to the edges labeled ${\bf b_i}, {\bf d_i}$ but with identical irreps on all other edges, which we refer to collectively as $\x = \{{\bf a_i}, {\bf c}, {\bf e_i}, \ldots\}$. We will refer to each such block of states as $B^{\x}$.
The values taken by each of the $D = m + n$ labels ${\bf b_i, d_i}$ are constrained by the fusion rules of these irreps with the irreps labeled ${\bf a_i},{\bf  e_i}$.  The number of states in such a block typically grows exponentially in $D$:
\begin{equation}
|B^{\x}| \sim  d_{{ a_1}} \ldots d_{ a_m} d_{ e_1} \ldots d_{ e_n} \sim   2^{\nu D},
\end{equation}
where $d_a$ is the (quantum) dimension of the irrep (anyon) ${\bf a}$.
The exponent $$\nu = \sum_a p_a \log_2 d_a,$$ where $p_a$ is the fraction of the irreps labeled ${\bf a_i},{\bf e_i}$ that are of type ${\bf a}$. In the infinite temperature ensemble of tree states, for edges sufficiently far from the bottom of the tree, the fraction of states where a given edge is labelled by $a$ is
\begin{equation}
p_a = \frac{d^2_a}{\sum_a d^2_a},
\label{eq:probdist}
\end{equation} 
independently of the starting representations~\cite{preskillLectureNotesQuantum2004}.

As both the operator $O^q_{r_1, r_2}$ and the tree states have been expressed as contractions of Clebsch-Gordan tensors, the matrix elements of $O^q_{r_1, r_2}$ can be expressed as well as a tensor network comprised purely of Clebsch-Gordan tensors. However, rather than performing a complicated tensor contraction, the properties of Clebsch-Gordan tensors allow for a vast simplification of this overlap calculation, reducing in the end to a simple product of $D+1$ scalar numbers with no tensor contractions. 
To derive our formula, it is helpful to consider the tensor contraction
\begin{center}
  \begin{tikzpicture}[level/.style={sibling distance=50mm/#1, level distance = 1cm},
                      level 1/.style={level distance = 1.cm},
                      level 2/.style={sibling distance=50mm},
                      level 5/.style={level distance = 1.cm},
                      level 6/.style={level distance = 1.cm},
                      edge from parent/.style={draw,postaction={on each segment={mid arrow=arrowstyle}}},
                      edge/.style={draw,postaction={on each segment={early arrow=arrowstyle}}},
                      lateredge/.style={draw,postaction={on each segment={later arrow=arrowstyle}}}
                      ]
    \usetikzlibrary{calc}
    \node[] at (-4, -2) {$O^q_{r_1, r_2} \ket{\Psi} =$};
    
    \def \ang{30}
    \node (c) {}
      child {node [circle,draw] (z) {}
        child {node [circle,draw] (a) {}
          child {node [rotate=-\ang] (b) {$\vdots$}
          }
          child {node [circle,draw] (g) {}
            child {node [rotate=-\ang] (s) {$\vdots$}
            }
            child {node [circle,draw] (i) {}
              child {node [rotate=-\ang] (f) {$\vdots$}
              }
              child {node [circle, draw] (h) {}
                child [grow=down] {node (o) {}}
              }
            }
          }
        }
        child {node [circle,draw] (j) {}
          child {node [circle,draw] (k) {}
            child {node [circle,draw] (m) {}
              child {node [circle, draw] (d) {}
                child [grow=down] {node (p) {}}
              }
              child {node (e) [rotate=\ang] {$\vdots$}
              }
            }
            child {node [rotate=\ang] (t) {$\vdots$}}
          }
        child {node [rotate=\ang] (l) {$\vdots$}
        }
      }
    };
    \path (h) -- (i) node [midway, below left=0.01cm and -0.1cm] {${\bf r_1}$};
    \path (d) -- (m) node [midway, below right=0.07cm  and -0.1cm] {${\bf r_2}$};
    \path (o) -- (h) node [midway, left=0.05cm] {${\bf r_1}$};
    \path (p) -- (d) node [midway, right=0.05cm] {${\bf r_2}$};
    \path (a) -- (z) node [midway, above=0.1cm] {${\bf b_m}$};
    \path (g) -- (a) node [midway, below left=-0.05cm and -0.15cm] {${\bf b_2}$};
    \path (i) -- (g) node [midway, below left=-0.05cm and -0.15cm] {${\bf b_1}$};
    \path (j) -- (z) node [midway, above=0.1cm] {${\bf d_n}$};
    \path (k) -- (j) node [midway, below right=-0.03cm and -0.1cm] {${\bf d_2}$};
    \path (m) -- (k) node [midway, below right=-0.03cm and -0.1cm] {${\bf d_1}$};
    \path (z) -- (c) node [midway, right=0.1cm] {${\bf c}$};
    \path (b) -- (a) node [midway, left=0.1cm]  {};
    \path (s) -- (g) node [midway, left=0.1cm]  {};
    \path (f) -- (i) node [midway, left=0.1cm]  {};
    \path (l) -- (j) node [midway, right=0.1cm] {};
    \path (t) -- (k) node [midway, right=0.1cm] {};
    \path (e) -- (m) node [midway, right=0.1cm] {};
    \node[right= 0.05cm of i]  (i2) {};
    \node[above right= 0.05cm of g]  (g2) {};
    \node[right= 0.1cm of a]  (a2) {};
    \node[below = -0.05cm of z]  (z2) {};
    \node[left = 0.1cm of j]  (j2) {};
    \node[above left = 0.05cm of k]  (k2) {};
    \node[left = 0.05cm of m]  (m2) {};
    \path[red, draw] (h.east) to[out=0, in=-70] (i2.center) -- (g2.center) -- (a2.center) -- (z2.center) -- (j2.center) -- (k2.center) -- (m2.center) to[out=250, in=180] (d.west);
    \draw[red, ->, line width=2,  >={stealth}, draw opacity=0] (z2) -- ($ 0.5*(a2) + 0.5*(z2)$) node [below right] {$q$};
  \end{tikzpicture}
\end{center}
Here, the edge with the irrep ${\bf q}$ has been drawn deformed to facilliate the next step in the derivation, 
which is to insert resolutions of the identity in terms of Clebsch-Gordan tensors
\begin{center}
\begin{tikzpicture}[on grid,
        ]
    \def \pd{0.6}
    \def \nd{0.5}
    \def \td{0.1}
    \node (center) {};
    \node[left = 2*\nd of center] (sigma) {$\nsum[2]\limits_{{\bf d'} \in {\bf d} \times {\bf q}}$};
    \node[above = \nd of center] (a) [circle, draw] {};
    \node[above left = \pd of a] (cl) {};
    \node[above right = \pd of a] (cr) {};
    \draw[mid arrow, line width=1] (cl.north)--(a) node [midway, left=\td] {${\bf d}$};
    \draw[red, mid arrow, line width=1] (cr.north)--(a) node [midway, right=\td] {${\bf q}$};
    \node[below = of a, circle, draw] (b) {};
    \node[below left = \pd of b] (dl) {};
    \node[below right = \pd of b] (dr) {};
    \draw[mid arrow, line width=1] (b)--(dl.south) node [midway, left=\td] {${\bf d}$};
    \draw[red, mid arrow, line width=1] (b)--(dr.south) node [midway, right=\td] {${\bf q}$};
    \draw[olive, mid arrow, line width=1] (a) -- node [midway, right=\td] {${\bf d'}$} (b);
    \node[left = (3.4*\nd) of center] (eq) {$=$};
    
    \node[left = 5*\nd of center] (center2) {};
    \node[above = (1.8*\nd) of center2] (a2) {};
    \node[below = (1.8*\nd) of center2] (b2) {};
    \node[left = (0.3*\pd) of a2] (cl2) {};
    \node[right = (0.3*\pd) of a2] (cr2) {};
    \node[left = (0.3*\pd) of b2] (dl2) {};
    \node[right = (0.3*\pd) of b2] (dr2) {};
    \draw[mid arrow, line width=1] (cl2.north) -- node [midway, left=\td] {${\bf d}$} (dl2.south);
    \draw[red, mid arrow, line width=1] (cr2.north) -- node [midway, right=\td] {${\bf q}$} (dr2.south);
\end{tikzpicture}
\end{center}
on each of the combined ${\bf q, d_i}$ edges and the equivalent for the ${\bf b}$ edges combined with the opposite orientation of ${\bf q}$. This results in a sum of diagrams of the following form:
\begin{center}
\begin{tikzpicture}[scale=1.2, line width=1,
  mid arrow/.style={postaction={decorate,decoration={
        markings,
        mark=at position .4 with {\arrowreversed[#1]{stealth}}
      }}}]
  \def \x{2.4}
  \def \sib{0.6}
  \node [circle, draw] (t) at (0, 0) {};
    \node [circle, draw] (t1) at ({-\x}, -1) {};
      \node [] (t11) at ({-\x-\sib}, -2) {};
      \node [circle, draw] (t12) at ({-\x+\sib}, -2) {};
        \node [] (t121) at ({-\x+0*\sib}, -3) {};
        \node [circle, draw] (t122) at ({-\x+2*\sib}, -3) {};
          \node [] (t1221) at ({-\x+1*\sib}, -4) {};
          \node [circle, draw] (t1222) at ({-\x+3*\sib}, -4) {};
    \node [circle, draw] (t2) at (\x, -1) {};
      \node [circle, draw] (t21) at ({\x-\sib}, -2) {};
        \node [circle, draw] (t211) at ({\x-2*\sib}, -3) {};
          \node [circle, draw] (t2111) at ({\x-3*\sib}, -4) {};
          \node [] (t2112) at ({\x-1*\sib}, -4) {};
        \node [] (t212) at ({\x+0*\sib}, -3) {};
      \node [] (t22) at ({\x+\sib}, -2) {};

  \draw[mid arrow, opacity=0.] (t) -- (t1);
    \draw[mid arrow] (t1) -- (t11);
    \draw[mid arrow, opacity=0.] (t1) -- (t12);
      \draw[mid arrow] (t12) -- (t121);
      \draw[mid arrow, opacity=0.] (t12) -- (t122);
        \draw[mid arrow] (t122) -- (t1221);
        \draw[mid arrow] (t122) -- (t1222);
  \draw[mid arrow, opacity=0.] (t) -- (t2);
    \draw[mid arrow, opacity=0.] (t2) -- (t21);
      \draw[mid arrow, opacity=0.] (t21) -- (t211);
        \draw[mid arrow] (t211) -- (t2111);
        \draw[mid arrow] (t211) -- (t2112);
      \draw[mid arrow] (t21) -- (t212);
    \draw[mid arrow] (t2) -- (t22);


  \def \up{0.8}
  \node (tup) at ($ (t) + (0, \up) $){};
  \draw[mid arrow] (tup) -- (t){};

  \def \down{0.8}
  \node [] (sL) at ($ (t1222) + (0, -\down) $){};
  \node [] (sR) at ($ (t2111) + (0, -\down) $){};

  \draw[mid arrow] (t1222) -- (sL){};
  \draw[mid arrow] (t2111) -- (sR){};


  \def \r{0.333};
  \def \del{0.1};
  \tikzstyle smallnode=[circle, inner sep=0pt, minimum size=1.5mm] {};

  \node [smallnode, draw] (s122) at ($ (t122) + (-\r*\sib+\del, \r+\del*\sib) $){};
  \node [smallnode, draw] (s12) at ($ (t12) + (-\r*\sib+\del, \r+\del*\sib) $){};
  \node [smallnode, draw] (q12) at ($ (t12) + (\r*\sib+\del, -\r+\del*\sib) $){};
  \node [smallnode, draw] (q1) at ($ (t1) + (\r*\sib+\del, -\r+\del*\sib) $){};
  \node [smallnode, draw] (s211) at ($ (t211) + (\r*\sib-\del, \r+\del*\sib) $){};
  \node [smallnode, draw] (s21) at ($ (t21) + (\r*\sib-\del, \r+\del*\sib) $){};
  \node [smallnode, draw] (q21) at ($ (t21) + (-\r*\sib-\del, -\r+\del*\sib) $){};
  \node [smallnode, draw] (q2) at ($ (t2) + (-\r*\sib-\del, -\r+\del*\sib) $){};

  \def \rr{0.25};
  \node [smallnode, draw] (s1) at ($ (t1) + (\x*\rr, \rr) $){};
  \node [smallnode, draw] (s2) at ($ (t2) + (-\x*\rr, \rr) $){};

  \node [smallnode, draw] (qL) at ($ (t) + (-\x*\rr, -\rr) $){};
  \node [smallnode, draw] (qR) at ($ (t) + (\x*\rr, -\rr) $){};

  \draw[mid arrow] (s1) -- (t1.60); 
  \draw[early arrow] (t1.-60) -- (q1); 
  \draw[early arrow] (t12.-60) -- (q12); 
  \draw[early arrow] (s12) -- (t12.120); 
  \draw[early arrow] (s122) -- (t122.120); 

  \draw[mid arrow] (t) -- (qL); 
  \draw[mid arrow] (t) -- (qR); 
  
  \draw[mid arrow] (s2) -- (t2.120);
  \draw[early arrow] (t2.240) -- (q2); 
  \draw[early arrow] (t21.240) -- (q21); 
  \draw[early arrow] (s21) -- (t21.60); 
  \draw[early arrow] (s211) -- (t211.60); 

  \draw [olive, early arrow] (q12) -- node[midway, below left] {${\bf b'_1}$} (s122);
  \draw [olive, early arrow] (q1) -- node[midway, below left] {${\bf b'_2}$} (s12);
  \draw [olive, mid arrow] (qL) -- node[midway, above] {${\bf b'_m}$} (s1);
  \draw [olive, early arrow] (q21) -- node[midway, below right] {${\bf d'_1}$} (s211);
  \draw [olive, early arrow] (q2) -- node[midway, below right] {${\bf d'_2}$} (s21);
  \draw [olive, mid arrow] (qR) --node[midway, above] {${\bf d'_n}$}  (s2);

  \draw [red, mid arrow reverse] (s122) to[out=0, in=0] (t1222);
  \draw [red, mid arrow] (s211) to[out=180, in=180] (t2111);

  \draw [red, mid arrow reverse] (s12) to[out=0, in=70] (q12);
  \draw [red, mid arrow reverse] (s1) to[out=-90, in=70] (q1);

  \draw [red, mid arrow] (s21) to[out=180, in=110] (q21);
  \draw [red, mid arrow] (s2) to[out=-90, in=110] (q2);

  \draw [red, mid arrow] (qL) to[out=-60, in=240] (qR);
\end{tikzpicture}
\end{center}
where the sum is over values of the irrep labels ${\bf b'_i} \in {\bf b_i} \times {\bf \bar{q}}$, ${\bf d'_i} \in {\bf d_i} \times {\bf q}$.
Finally, we can simplify each of these diagrams into a single tree state by substituting the following identity of Clebsch-Gordan tensors:
\begin{center}
\begin{tikzpicture}[line width=1]
  \def \sib{0.8}
  \def \up{1.2}
  \def \r{0.5};
  \def \rr{0.6};
  \def \del{0.};
  \tikzstyle smallnode=[circle, inner sep=0pt, minimum size=2mm] {};

  \node [circle, draw] (t) at (0, 0) {};
  \node [] (t1) at ({-\sib}, -1) {};
  \node [] (t2) at ({\sib}, -1) {};
  \draw[mid arrow] (t) -- node[midway, right] {$e$} (t2);
  \node (tup) at ($ (t) + (0, \up) $){};
  \node [smallnode, draw] (q) at ($ (t) + (-\r*\sib+\del, -\r-\del*\sib) $){};
  \node [smallnode, draw] (s) at ($ (t) + (0, \rr) $){};

  \draw [mid arrow] (s) -- node[midway, right] {$d$} (t);
  \draw [mid arrow] (t) -- node[midway, below right] {$b$}(q);
  \draw [red, mid arrow] (s) to[out=180, in=140] node[midway, left] {$q$} (q);
  \draw [olive, mid arrow] (tup) -- node[midway, right] {$d'$} (s);
  \draw [olive, early arrow] (q) -- node[midway, below] {$b'$} (t1);
  \node[] at (1.5, 0) {$= \left(F^{q b' e}_{d}\right)_{b d'}$};

  \def \y{3};
  \node [circle, draw] (x) at (\y, 0) {};
  \node [] (x1) at ({\y -\sib}, -1) {};
  \node [] (x2) at ({\y + \sib}, -1) {};
  \node [] (xup) at ({\y}, \up) {};
  \draw [olive, mid arrow] (x) -- node[midway, below right] {$b'$} (x1);
  \draw [mid arrow] (x) -- node[midway, right] {$e$} (x2);
  \draw [olive, mid arrow] (xup) --node[midway, right] {$d'$} (x);
\end{tikzpicture}
\end{center}
along the right branch of the diagram and a similar one along the left branch. After these substitutions, each diagram in the sum is a single tree state with a coefficient that consists of $m+n+1$ $F$-symbols.
The $F$-symbols are a group theoretic factor depending on $6$-irrep labels defined via a contraction of $4$ Clebsch-Gordan tensors. From this we can read off the matrix element of the operator $O^q_{r_1, r_2}$ between any two tree states: 
\begin{multline}
\bra{\Psi'} O^{(q)}_{r_1, r_2} \ket{\Psi} = 
\delta_{\x \x'} \left(\prod\limits_{i=1}^{m} \left(F^{a_i b_{i-1} q}_{b'_i} \right)_{b_i b'_{i-1}}\right) \\
\left(F^{b_m q d'_n}_c\right)^{\dagger}_{d_n b'_m} 
\left(\prod\limits_{j=1}^{n} \left(F^{q d'_{j-1} e_{j}}_{d_j} \right)_{d_{j-1} d'_{j}}\right),
\label{eq:matrixelement}
\end{multline}
where the unprimed labels ${\bf a_i}, {\bf b_i}, {\bf c}, {\bf d_i}, {\bf e_i}$ refer to the irrep labels of $\ket{\Psi}$, the corresponding primed labels to $\ket{\Psi'}$, and ${\bf b_0} = {\bf b_0'} = {\bf r_1}, {\bf f_0} = {\bf f_0'} = {\bf r_2}$.
The same formula works for anyonic tree states, which are defined via the $F$-symbols without the underlying Clebsch-Gordan tensors.

Via Eq.~\ref{eq:matrixelement}, we have an analytic handle on precisely which matrix elements are non-zero and how big they are.
We see immediately that the connected tree states are not all of $B^{\x}$ but only those with irrep labels ${\bf b'_i} \in {\bf b_i} \times {\bf \bar{q}}$, ${\bf d'_i} \in {\bf d_i} \times {\bf q}$. This is a generalization of the selection rule described in Ref.~\onlinecite{protopopovEffectSUSymmetry2017} for \sutwo tree symmetric tree states.
In that example, the perturbing operator is $V = \vec{S}_i \cdot \vec{S}_{i+1}$, ${\bf q}$ is the spin-1 irrep, and tree states with a spin value of $S$ on a tree edge are only connected to tree states with $S' \in S \times {\bf 1} = \{S -1, S, S + 1\}$ on that edge.
These selection rules generally allow for multiple possible labels for each of the $D$ labels ${\bf b'}, {\bf d'}$ as long as $d_q>1$. The typical number of non-zero matrix elements from each tree state is thus asymptotically exponential in $D$:
\begin{equation}
N^q \sim 2^{\alpha_q D}.
\label{eq:expN}
\end{equation}
The exponent $\alpha_q$ depends only on the group theoretic data and ${\bf q}$. We compute it for various examples in the following sections using Eq.~\ref{eq:matrixelement}.
As at most all of the states in $B^{\x}$ can be connected, we have the constraint that $0 \leq \alpha_q  \leq \nu$.

The values of the matrix elements are products of $D$ $F$-symbols, each of which is less than $1$ in absolute value, and thus the size of the non-zero matrix elements is asymptotically exponentially decaying in $D$:
\begin{equation}
|V^q_{ab}| \sim 2^{- \beta_q D}.
\label{eq:expV}
\end{equation}
Again, the exponent only depends on $q$, and we will compute it for various examples in the following sections.

The energy denominators $E_a - E_b$ between connected tree states, which enter our criterion for resonances, are challenging to describe. First, we consider the simplest scenario for the behavior of these quantities, which suggests that resonances should be increasingly common as $D$ increases. The examination of the accuracy of this scenario is postponed to Sec.~\ref{fig:resonantcounting}.
Suppose that the $2^{\alpha_q D}$ energy levels of the connected tree states are distributed uniformly and randomly in an energy window, with a typical level spacing $2^{-\alpha_q D}$. Then nearby states in the energy spectrum will have
$$\mathcal{G}^V_{n, n+1} \sim \log 2^{(\alpha_q - \beta_q) D} \sim \log 2 \cdot \gamma_q D,$$
with $\gamma_q =  \alpha_q - \beta_q$. Under the same assumption,
the expected number of resonant connections from a given state is 
\begin{equation}
N_{R} \sim \lambda 2^{(\alpha_q - \beta_q)D},
\label{eq:NR}
\end{equation}
which is the expected number of connected states that lie in an energy window of size $\lambda V$.

For every tree shape, there are some bonds for which the connecting path between them reaches near the top of the tree, which requires a length of path logarithmic in the number of sites $L$.
Thus the exponential scaling of the resonant connections in Eq.~\eqref{eq:NR} in $D$ translates to a power law scaling of $L$ for such worst case bonds. Specifically, let $D_{i, i+1}$ be the depth $D$ relevant for an operator at the sites $i, i+1$. Ref.~\onlinecite{protopopovNonAbelianSymmetriesDisorder2019} shows that, for a specific model of random tree shapes where the locations of spin fusions are independent and uniform, $D_{i, i+1}$ are distributed according to
\begin{equation*}
p(D) = \frac12 \left( \frac23 \right)^D, \; \; D\ge1.
\end{equation*}
The typical maximum value of $D$ for a system of $L$ sites can be obtained from the condition $p(D_{\text{max}}) \sim 1/L$, giving
\begin{equation}
D_{\text{max}} \sim \frac{\log L}{\log \frac32}.
\end{equation}

Among two-site operators, the most destabilizing perturbation is $O^q$ at the cut of the tree with the largest $D$, and with $q$ chosen among operators to give the one most likely to create resonances, i.e. with the largest $\gamma_q$.
For this perturbation, Eq.~\ref{eq:beta} becomes
$$\mathcal{G}^V_{n, n+1} \sim \gamma' \log L$$ 
with
$$\gamma'  = \frac{\log 2}{\log \frac32} \max_q \gamma_q.$$
The tree eigenstates are unstable if any $\gamma_q > 0$.

This model of tree shapes is not necessarily the one realized by RSRG-X for a given system, but depths of random trees are generally logarithmic in the number of leaves. We parametrize this unknown with a constant $\rho$, so that 
\begin{equation}
D_{\text{max}} \sim \rho \log_2 L.
\label{eq:rho}
\end{equation}
Then instead $\gamma' = \max_q \rho \gamma_q$ and the number and size of matrix elements scale as power-laws $N^q \sim L^{\rho \alpha_q}$ and $V^q \sim L^{-\rho \beta_q}$. The value of $\rho$ does not affect whether tree states are stable but does affect the relevant thermalization length and time scales.

The exact matrix element formula Eq.~\eqref{eq:matrixelement} thus gives us a quantitative window into the stability of a broad swath of non-Abelian QCGs without using much information about the microscopics.
The rest of this paper will be spent evaluating these stability criteria and confirming the picture outlined in this section.
In Section~\ref{sec:numericmatrixelements} we will support our results regarding the size and number of matrix elements by randomly sampling tree states and numerically evaluating Eq.~\ref{eq:matrixelement} for particular non-Abelian groups and anyon theories. As these matrix elements are relatively easy to compute, we can access large systems of up to $L=2^{15}$ spins, making the exponential dependence on $D$ clear. In Section~\ref{sec:randommatrixproducts}, we discuss an even easier way to compute $\alpha_q$ and $\beta_q$ directly in the thermodynamic limit without the need to sample tree states. For some cases, our method even yields analytic formulas. 

Finally, we remind the reader that the biggest limitation of this Section is the modeling of energy denominators.
To check whether the simplified picture regarding the statistics of energy denominators described above yields the correct asymptotic scaling of the number of resonances, we do an exact counting of resonances in Section~\ref{sec:directcomputation} finding strong supporting evidence for the scenario presented here.

\section{Size and number of matrix elements}
\label{sec:sizeandnumber}
\subsection{Numerical computation}
\label{sec:numericmatrixelements}

To verify that the scaling of the number of non-zero matrix elements and their values are indeed captured by Eqs.~\eqref{eq:expN}-\eqref{eq:expV}, we now explicitly compute exact matrix elements and generate their statistics.
To do so, we generated random tree states with the following procedure, 
starting with an initial spin chain consisting of spins transforming as a non-Abelian irreducible representation ${\bf r}$. We then repeatedly choose a pair of neighboring spins at random to fuse. The fusion outcome is chosen using the infinite temperature ensemble 
$$
p(a, b \to c) = \frac{d_c}{d_a d_b} \text{ for } {\bf c} \in {\bf a} \times {\bf b}, \text{ else }0.
$$

\begin{figure*}
    \centering
    \includegraphics[width=0.8\linewidth]{{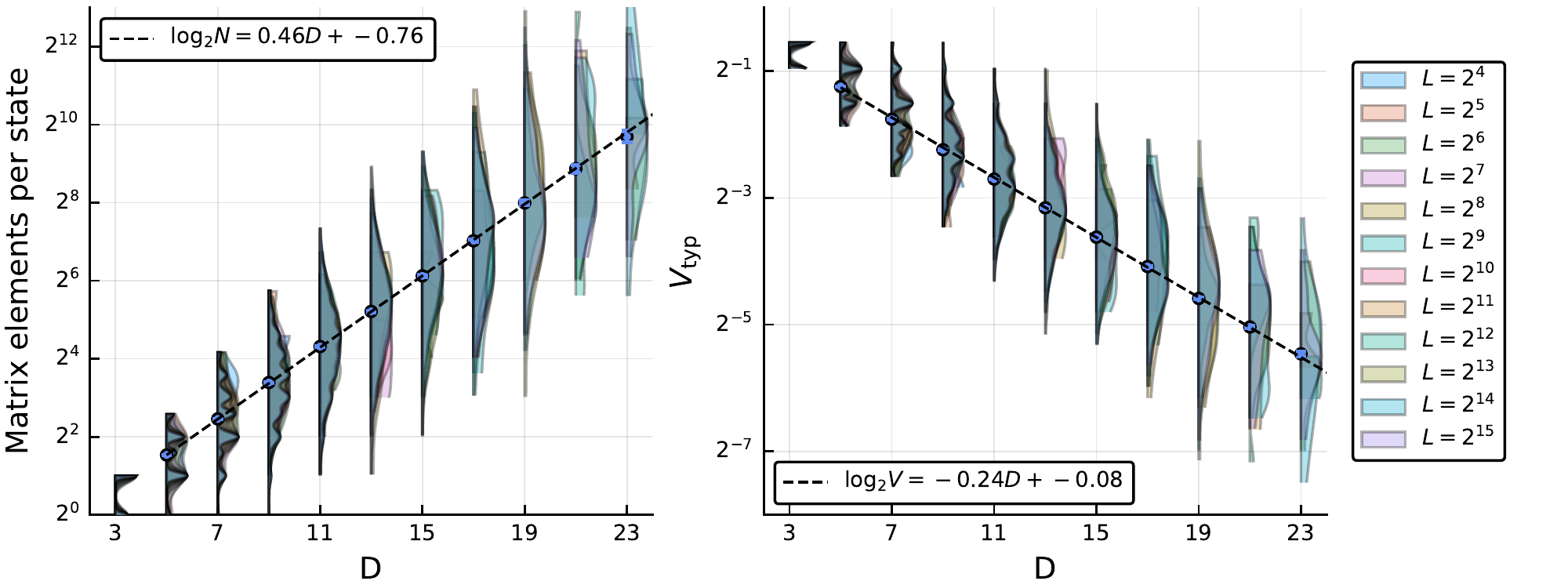}}
    \includegraphics[width=0.8\linewidth]{{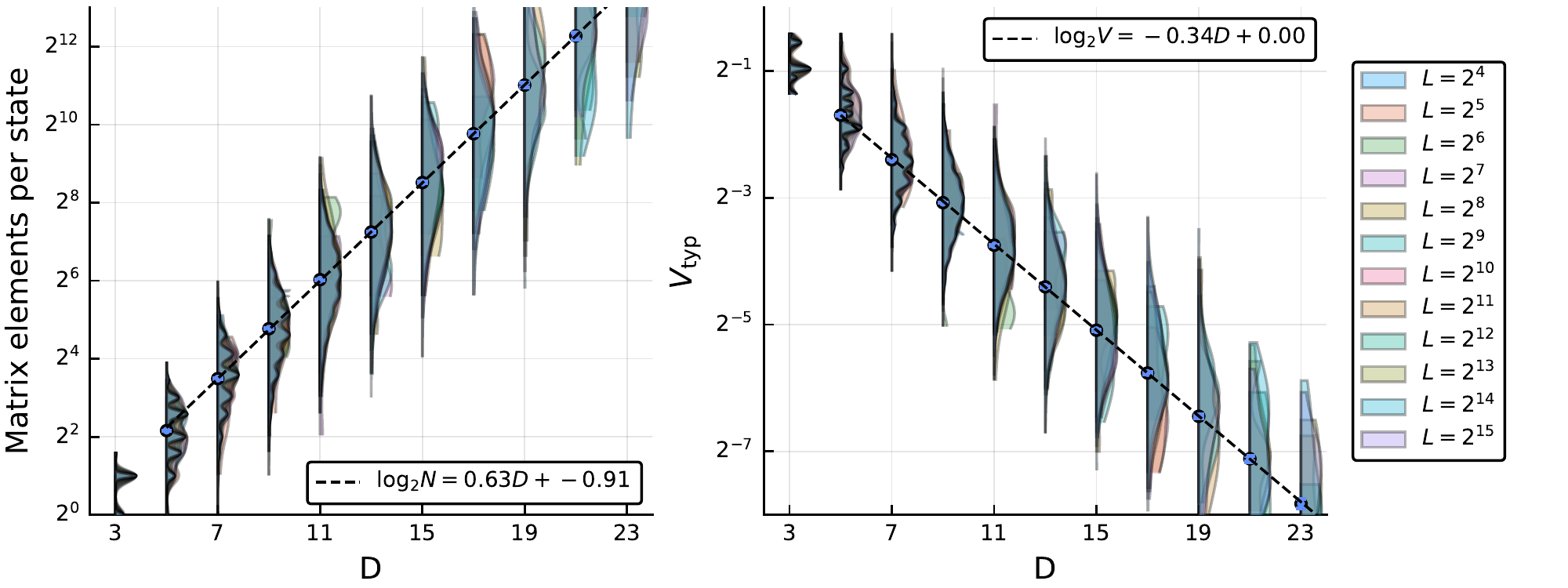}}
    \caption{Number (left) and typical size (right) of matrix elements in the basis of tree states on $L$ spins, sampled from random tree states. The matrix element distribution asymptotically depends only on $D$. The exponents closely match the Lyapunov calculation results for $\alpha$ (left) and $\beta$ (right) shown in Table~\ref{table:groupanalytic}. \\
    (top) Matrix elements of the $D_3$ invariant operator $O^{\mathbf{2}}_{\mathbf{2}\mathbf{2}}$.
    (bottom) Matrix elements of the \sutwok{4} operator $O^{\mathbf{1}}_{\mathbf{\frac12},\mathbf{\frac12}}$.
    }
    \label{fig:numericmatrixelements}
\end{figure*}

With each tree state $\ket{\Psi}$, we generate \emph{all} non-zero matrix elements connecting that state to other states (using Eq.~\ref{eq:matrixelement}) for each two-site perturbation $O^{q}_{r, r}$ and at every possible position of the operator at neighboring sites $i, i+1$.
This can be done efficiently by iterating through each possibility ${\bf b_i'} \in {\bf b_i} \times {\bf q}$, ${\bf d_i'} \in {\bf d_i} \times \bar{{\bf q}}$ for the irrep labels of a connected state $\ket{\Psi'}$ and checking whether the resulting tree state is valid. 
Using this procedure, we can reach large system sizes of up to $L=2^{15}$ spins, which we find is more than sufficient to obtain the asymptotic scaling.
For each state sampled, we tabulate the number of non-zero matrix elements involving that state and the typical size $\exp \braket{ \log \left| V \right|}$ of the non-zero matrix elements $\braket{\Psi' | O^{q}_{r, r} | \Psi}$. 
Results for two examples, the \Dthree symmetric (Dihedral group) spin chain and the \sutwok{4} anyon chain, are shown in Fig.~\ref{fig:numericmatrixelements}. 
We see that the properties of the matrix elements are as described in the previous section: the typical size and number of matrix elements of a local perturbation scale exponentially in $D$. Additionally we see that these quantities have broad, lognormal-like distributions.

For \Dthree, there are three irreducible representations, commonly referred to as the trivial irrep, the sign irrep, and the fundamental irrep. The dimensions of these irreps are $1$, $1$, and $2$ respectively --- for this reason the fundamental irrep is denoted as $\mathbf{2}$.
We will take the initial configuration of spins for our \Dthree symmetric spin chain to be $L$ spins that transform as ${\bf r} =\mathbf{2}$. The perturbing operator we use is $O^{\mathbf{2}}_{\mathbf{2},\mathbf{2}}$. This is the only choice for ${\bf q}$ that can lead to $\alpha_q > 0$, as taking ${\bf q}$ to be an Abelian one-dimensional irrep gives at most one connected tree state. Plotting the typical number of connected states versus $D$ and fitting gives $\alpha_{\mathbf{2}} \approx 0.46$. Plotting the typical size of the matrix elements versus $D$ and fittings gives $\beta_2 \approx 0.24$. In Section~\ref{sec:randommatrixproducts} we find exact expressions for these exponents from an analytic calculation, which match the numerically extracted values within the margin of error of our fitting.

For our second example we use \sutwok{4} anyons. In this case, the anyon labels can take $5$ types, conventially called $\mathbf{0}, \mathbf{\frac12}, \mathbf{1}, \mathbf{\frac32}, \mathbf{2}$, with dimensions $1, \sqrt{3}, 2, \sqrt{3}, 1$ respectively. 
In this case the basis of perturbing operators $O^q_{r, r}$ is spanned by ${\bf q} =\mathbf{0}, \mathbf{1}, \mathbf{2}$, as these are the only anyon types $q$ satisfying $N^r_{q,r} > 0$. Of these, only ${\bf q}=\mathbf{1}$ also has $d_q >1$.
We thus take the starting configuration to be a chain of ${\bf r}=\mathbf{\frac12}$ anyons and the pertubing operator to be $O^{\mathbf{1}}_{\mathbf{\frac12},\mathbf{\frac12}}$.
As above, we sample tree-states and find clear exponential scaling with exponents $\alpha_{\mathbf{1}} \approx 0.63$ and $\beta_{\mathbf{1}} \approx 0.34$ extracted from the fit in Fig.~\ref{fig:numericmatrixelements}.

In each case, $\alpha > \beta$ and so under the hypothesis of the previous section we find that $\gamma > 0$ and the perturbative instability criteria are satisfied.
We also computed the number and size of matrix elements for other examples of non-Abelian symmetry groups and anyon theories. Each asymptotically follows the predictions of Eqs.~\ref{eq:alpha}-\ref{eq:beta} with exponents matching those derived in the following, in Section~\ref{sec:randommatrixproducts} and displayed in Table~\ref{table:groupanalytic}.

\subsection{Random matrix product calculation}
\label{sec:randommatrixproducts}
An appealing feature of Eq.~\ref{eq:matrixelement} is that the indices on the factors align as if the matrix element were one term in the expansion of a product of matrices. We exploit this structure below to compute the asymptotic scaling properties of the matrix elements.
In particular, we can express the following three quantities as products of 
matrices:
\begin{itemize}
    \item $N^{\x}$, the number of states in $B^\x$,
    \item $M^{\x, q}$, the total number of non-zero matrix elements of $O^q$ between pairs of states in $B^\x$,
    \item $F^{\x, q}$, the sum of the absolute value of these matrix elements.
\end{itemize}
Given these quantities, we can extract the mean number of non-zero matrix elements per state $M^{\x, q}/N^\x$ and the mean size of the non-zero matrix elements $F^{\x, q}/M^{\x, q}$.

We can count the number of states in $B^{\x}$ by the following sum, where the unfixed labels $b, d$ vary over all unconstrained possibilities and the summand is $1$ if all nodes in the tree respect the fusion rules or $0$ otherwise:
\begin{equation}
\begin{split}
N^{\x} =& \sum \limits_{b, d} N^{b_1}_{a_1 r_1} \cdots N^{b_m}_{a_m b_{m-1}} N^{c}_{b_m d_{n}} N^{d_n}_{e_n d_{n-1}} \cdots N^{d_1}_{e_1 r_2} \\
 =& \left(n^{a_1} \cdots n^{a_m} \tilde{n}^{c} n^{e_n T} \cdots n^{e_1 T} \right)_{r_1 r_2},
\end{split}
\label{eq:matrixproductN}
\end{equation}
where $n^a$ are matrices with matrix elements $\left(n^a\right)_{bc} = N_{ab}^c$ and $\tilde{n}^a$ are matrices with matrix elements $\left(\tilde{n}^a\right)_{bc} = N_{bc}^a.$
We can compute $M^{\x, q}$ analogously, but now with a double sum over pairs of states with labels $b, d$ and $b', d'$, and the summand being $1$ if the corresponding matrix element is non-zero. This turns into an analogous matrix product:
\begin{equation}
\begin{split}
M^{\x, q} =& \left(m^{a_1} \cdots m^{a_m} \tilde{m}^{c} \bar{m}^{e_n} \cdots \bar{m}^{e_1} \right)_{(r_1 r_1) (r_2 r_2)},
\end{split}
\label{eq:matrixproductM}
\end{equation}
where $m$, $\tilde{m}$, and $\bar{m}$ are matrices with matrix elements
$$
m^{a, q}_{(bb')(cc')} = 1 \text{ if } \left(F^{a b q}_{c'} \right)_{c b'} \neq 0 \text{ else } 0,
$$
$$
\tilde{m}^{c, q}_{(bb')(dd')} = 1 \text{ if } \left(F^{b q d'}_{c} \right)^{\dagger}_{d b'} \neq 0 \text{ else } 0,
$$
$$
\text{and }\bar{m}^{e, q}_{(bb')(cc')} = 1 \text{ if } \left(F^{q c' e}_{b} \right)_{c b'} \neq 0 \text{ else } 0.
$$
Finally, $F^{\x, q}$ is the same 
\begin{equation}
\begin{split}
F^{\x, q} =& \left(f^{a_1} \cdots f^{a_m} \tilde{f}^{c} \bar{f}^{e_n} \cdots \bar{f}^{e_1} \right)_{(r_1 r_1) (r_2 r_2)},
\end{split}
\label{eq:matrixproductF}
\end{equation}
with 
$$
f^{a, q}_{(bb')(cc')} = \left|\left(F^{a b q}_{c'} \right)_{c b'} \right|,
$$
$$
\tilde{f}^{c, q}_{(bb')(dd')} = \left|\left(F^{b q d'}_{c} \right)^{\dagger}_{d b'} \right|,
$$
$$
\text{and }\bar{f}^{e, q}_{(bb')(cc')} = \left| \left(F^{q c' e}_{b} \right)_{c b'} \right|.
$$
The labels $a_i$, $e_i$
which determine the factors in the matrix product in Eqs.~(\ref{eq:matrixproductN})-(\ref{eq:matrixproductF}) are independent, as each is the result of fusing disjoint sets of initial irreps/anyons. These labels can be described as being randomly sampled from the equilibrium probability distribution $p_a$ from Eq.~\ref{eq:probdist}.

Therefore, the typical behavior of $N^\x, M^{\x,q},$ and $F^{\x,q}$ is governed solely by the growth of a product of $D+1$ random matrices, each sampled from a fixed set of matrices --- one for each anyon or irrep type --- with the given probability distribution $p_a$.
In Eqs.~(\ref{eq:matrixproductN})-(\ref{eq:matrixproductF}), there are three distinct sets of matrices used --- one for the first $m$ factors, another for the middle factor, and a third for the last $n$ factors. This happens because of the direction of ingoing and outgoing arrows along the left and right side of the local geometry of the tree in Eq.~\ref{eq:matrixelement}. As there is no physical distinction between the left and right side of the tree, the scaling properties are the same when using either set of matrices.

Such random matrix products generically grow exponentially, with the asymptotic growth controlled by an exponent known as the leading Lyapunov exponent. We find that this is true as well for the quantities in Eqs.~(\ref{eq:matrixproductN})-(\ref{eq:matrixproductF}).
Let the three growth exponents be $\nu, \mu, \xi,$ so that 
\begin{align}
N^{\x} &\sim 2^{\nu D} \nonumber\\ 
M^{\x, q} &\sim 2^{\mu_q D} \\
F^{\x, q} &\sim 2^{\xi_q D}.\nonumber
\end{align}
From these exponents, we can surmise that the mean number of non-zero matrix elements per state is 
\begin{equation}M^{\x, q}/N^{\x} \sim 2^{\alpha_q D}, \quad \alpha_q=\mu_q - \nu. \label{eq:alpha}\end{equation}
Similarly the mean size of a non-zero matrix element is 
\begin{equation}F^{\x, q}/M^{\x, q} \sim 2^{-\beta_q D}, \quad \beta_q=\mu_q - \xi_q. \label{eq:beta}\end{equation}
We are particularly interested in the scaling of \emph{off-diagonal} matrix elements, while these formulas include diagonal matrix elements as well.
Eq.~\ref{eq:matrixelement} behaves the same for diagonal and off-diagonal matrix elements --- each is a product of the same number of $F$-symbols --- and so we do not expect much difference in their magnitudes. As there are many fewer diagonal matrix elements than off-diagonal, the diagonal matrix elements only give a subleading contribution to the sum of matrix elements $F^{\x}$ and the number of matrix elements $M^{\x}$, and thus do not effect the scaling of these quantities.
This can be easily confirmed by modifying Eqs.~(\ref{eq:matrixproductM},\ref{eq:matrixproductF}) to compute the number and sum of only the diagonal matrix elements.

\begin{table}[t!]
\centering
\renewcommand{\arraystretch}{1.5}
\setlength{\tabcolsep}{0.3em}
\begin{tabular}{ll||rrr|rr|r}\hline
\iA &  $q$ & $\nu$ & $\mu$ & $\xi$ &$\alpha$&$\beta$&$\gamma$\\ \hline \hline
\Dthree & $\mathbf{2}$ & 0.667 & 1.130 & 0.893 & 0.463 & 0.236 & 0.227 \\
\Dk{5} & $\mathbf{2_j}$ & 0.800 & 1.159 & 0.970 & 0.359 & 0.189 & 0.170 \\
\Dk{7} & $\mathbf{2_j}$ & 0.857 & 1.146 & 0.991 & 0.288 & 0.155 & 0.133 \\
\hline
\sutwok{2}/\Ising & $\mathbf{\frac12}\text{/}\mathbf{\sigma}$ & 0.250   & 0.500 & 0.375 & 0.250 & 0.125 & 0.125 \\
\hline
\sutwok{2}/\Ising & $\mathbf{1}\text{/}\mathbf{\psi}$ & 0.250   & 0.250 & 0.250 & 0.000 & 0.000 & 0.000 \\
\sutwok{3} & $\mathbf{1}$ & 0.502 & 0.920 & 0.702 & 0.418 & 0.218 & 0.200 \\
\sutwok{4} & $\mathbf{1}$ & 0.730 & 1.356 & 1.031 & 0.627 & 0.326 & 0.301 \\
\sutwok{5} & $\mathbf{1}$ & 0.931 & 1.750 & 1.307 & 0.819 & 0.444 & 0.375 \\
\sutwok{6} & $\mathbf{1}$ & 1.111 & 2.016 & 1.530 & 0.905 & 0.485 & 0.420 \\
\hline 
\sothreek{3}\text{/}\Fib  & $\mathbf{1}\text{/}\mathbf{\tau}$ & 0.502 & 0.920 & 0.702 & 0.418 & 0.218 & 0.200 \\
\sothreek{4} & $\mathbf{1}$ & 0.667 & 1.130 & 0.893 & 0.463 & 0.236 & 0.227 \\
\sothreek{5} & $\mathbf{1}$ & 0.931 & 1.750 & 1.307 & 0.819 & 0.444 & 0.375 \\
\sothreek{6} & $\mathbf{1}$ & 1.085 & 1.982 & 1.493 & 0.897 & 0.489 & 0.407 \\
\hline \hline
\end{tabular}
 \caption{Tabulated exponents computed using Lyupanov transfer matrices for dihedral groups $D_{2k+1}$ and for the anyon theories \sutwok{k} and \sothreek{k}. Note the equivalence of \sothreek{4} and \Dk{3}, and equal exponents for \sothreek{k} and \sutwok{k} for odd $k$.}
\label{table:groupanalytic}
\end{table}

\begin{figure}
\centering
\includegraphics[width=0.9\linewidth]{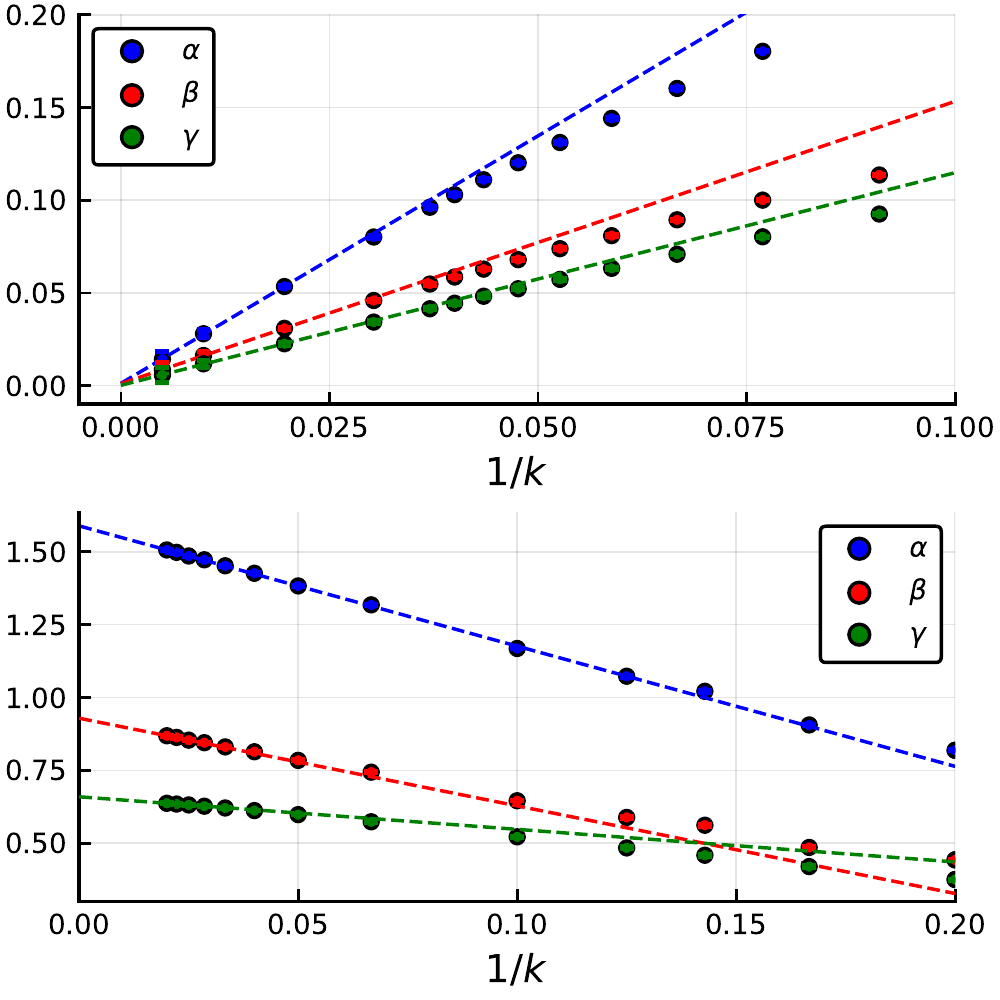}
\caption{(top) For the dihedral groups $D_{k}$: $\alpha$, $\beta$, $\gamma$ $ \to 0$ as $k \to \infty$.
(bottom) For \sutwok{k} anyon chains with $q=\mathbf{1}$: $\alpha \to \log_2 3, \beta \to 0.93, \gamma \to 0.66$ as $k \to \infty$.}
\label{fig:dihedral}
\end{figure}

These random matrix products therefore give us an alternate way to compute the exponents and evaluate the stability criteria in Section~\ref{sec:resonant_processes}.
We compute these exponents using a straightforward Monte Carlo method, measuring the growth of the size of a random initial vector while multiplying it with randomly sampled matrices from the set. Millions of samples are needed to converge the estimate of the exponent; however, as each matrix-vector multiplication is of a small fixed size $d^2$ where $d$ is the number of irrep/anyon types, only a few minutes of computation are needed to obtain the exponents to $10^{-3}$ accuracy, directly in the thermodynamic limit. 
We have computed the Lyapunov exponents numerically for a number of non-Abelian symmetry groups and anyon systems and summarized the results in Table~\ref{table:groupanalytic} and Fig.~\ref{fig:dihedral}. 

As an additional check on our Monte Carlo method, we see that our estimated Lyapunov exponents match exact values in several cases where they are available.
For all cases, our numerically computed exponents of $\nu$ match the value 
$$
\nu = \frac{\sum_a d_a^2 \log_2 d_a}{\sum_a d_a^2},
$$
from Section~\ref{sec:resonant_processes}.
In Appendix~\ref{sec:exact}, we show that the exponents for \Dthree are exactly
\begin{align*}
\nu &= \frac23, \quad 
\mu = \frac23 \log_2 \left( 1 + \sqrt{5} \right) \approx 1.129496, \\
\text{ and } \xi &= \frac23 \log_2 \left(\frac{1 + \sqrt{1+4\sqrt{2}}}{\sqrt{2}} \right) \approx 0.893330.
\end{align*}
Similarly, the exponents for Fibonacci anyon chains are
\begin{gather*}
\nu = \frac{1+\varphi}{2+\varphi} \log_2 \varphi, \quad 
\mu = \frac{1+\varphi}{2+\varphi} \log_2 \left( 1 + \sqrt{2} \right), \text{ and } \\\xi =
\frac{1+\varphi}{2+\varphi} \log_2 \frac{\varphi^{\frac12} + 1 + \sqrt{1 - 2 \varphi^{\frac12} + \varphi + 8\varphi^{\frac32}}}{2 \varphi}.
\end{gather*}
These values match the results in Table~\ref{table:groupanalytic}. 

Our Monte Carlo estimates for the exponents $\nu$, $\mu$, and $\xi$ and the consequent values for $\alpha$, $\beta$, and $\gamma$ are shown in Table~\ref{table:groupanalytic} for the dihedral groups $D_{2k+1}$, for non-Abelian \sutwok{k} anyons, and for \sothreek{k} anyons (which correspond to restricting \sutwok{k} anyons to only the subset with integer spins). In all cases considered except one, $\gamma > 0$; the odd case is that of the Ising anyon chain with $\gamma=0$, discussed further below. These families include several previously studied potential QCG phases. Refs.~\onlinecite{prakashEigenstatePhasesFinite2017,friedmanLocalizationprotectedOrderSpin2018} considered spin chains with \Dthree symmetry, in which the potential QCG phase occurs between spin glass phases which spontaneously break the symmetry down to an Abelian subgroup. This phase diagram in particular contains a self-dual point that is known to not break the symmetry. Our finding of $\gamma>0$ shows that the QCG is unstable, but does not a priori indicate whether the leading instability is towards spontanous symmetry breaking or thermalization. The QCG region either shrinks to a fine-tuned critical point (at the self-dual point) or is instead replaced by an intervening ergodic phase. Anyon chains, on the other hand, cannot by construction spontaneously break the symmetry --- so any instability can only be towards thermalization~\footnote{It is worth nothing that quantum critical wave functions can be dramatically affected by local perturbations in clean systems, leading to an ``orthogonality catastrophe''~\cite{PhysRevLett.18.1049}. In principle, one could imagine a similar scenario in excited states (though it is hard to imagine it would be robust against thermalization), which would provide another option with an instability towards another set of tree states. However, infinite randomness fixed points turn out to be stable to local perturbations at zero temperature~\cite{PhysRevB.92.054203}, and do not suffer from an Anderson orthogonality catastrophe, so this possibility is ruled out. }. The QCGs of \sutwok{k} anyon chains were previously considered in Ref.~\onlinecite{vasseurQuantumCriticalityHot2015} and our finding of $\gamma>0$ suggests asymptotic thermalization for all of these chains. 

For Ising anyon chains, there are three anyon types, traditionally denoted $\mathbf{1}, \mathbf{\sigma}, \mathbf{\psi}$, with dimensions $1, \sqrt{2}, 1$ respectively. The only operator with more than one matrix element should then be $O^{q}_{ab}$ with ${\rm q}=\sigma$. However, the fusion rules of the theory do not allow for an operator $O^{\sigma}_{\sigma\sigma}$. For a chain of $\sigma$ anyons, there are thus only Abelian operators such as $O^{\psi}_{\sigma\sigma}$ and the corresponding exponents $\alpha, \beta, \gamma$ are $0$. For this reason, our analysis does not produce a conclusion about the stability of the QCG in Ising anyon chains or similarly, parafermion chains. An alternate viewpoint is that these anyon chains are dual to crtical points in Abelian spin chains with $\mathbb{Z}_2$ symmetry ($\mathbb{Z}_q$ for parafermions) rather than non-Abelian spin chains and thus outside the scope of this paper. We note that two very recent papers claimed --- largely based on numerical evidence from exact diagonalization --- that the phase transition between MBL phases in Ising ($\mathbb{Z}_2$) chains are {\em perturbatively} unstable against thermalization~\cite{2020arXiv200809113M, 2020arXiv200808585S}. It would be interesting to generalize our approach to deal with Abelian critical MBL states to explain this instability analytically.
To realize an operator with $d_q > 0$ with Ising anyons, one can consider a chain with each site represented by $\mathbf{1} + \sigma$, where there is a local operator $O^{\sigma}_{ab}$ with $a=b=\mathbf{1} + \sigma$. The interpretation of such an operator could be that it corresponds to the hopping of a $\sigma$ anyon in a chain of itinerant $\sigma$ anyons~\cite{Poilblanc2012}. In this case, we again find $\gamma>0$.

We also examined the limiting behavior of two our families of non-Abelian chains as the group becomes large.
For dihedral group symmetric chains with odd $k>3$, there are two one-dimensional irreps and $(k-1)/2$ two-dimensional irreps labeled $\mathbf{2}_j$, $j \in [1, \ldots, \frac{k-1}{2}]$. The Lyapunov exponents are the same for each choice of $q = \mathbf{2}_j$. While $\gamma>0$ for each $k$, in the limit $k \to \infty$, $\alpha, \beta$, and $\gamma$ approach $0$, as shown in Fig.~\ref{fig:dihedral}. Intuitively, the reason for this limiting behavior is that all $F$-symbols of $D_{2k+1}$ are $1$ except those involving the two one-dimensional irreps. These irreps become a vanishingly small proportion of tree labels as $k \to \infty$ in the equilibrium distribution Eq.~\ref{eq:probdist}.
In the $k \to \infty$  limit, the representation theory and $F$-symbols of $D_{2k+1} = {\mathbb Z}_{2k+1} \rtimes  {\mathbb Z}_{2}$ approach those of the group $\Otwo = {\rm U}(1)  \rtimes  {\mathbb Z}_{2}$, which has an infinite number of irreps.
However, starting from spins that transform as some fixed irrep, say $\mathbf{2}_1$, the equilibrium distribution of irreps won't be reached until $L \gtrsim k$ --- and for $\Otwo$ tree states, it will never be reached.
This suggests that our approach breaks down and that the thermodynamic limit and the $k \to \infty$ limit do not commute. 

For \sutwok{k} chains, there are $k+1$ anyon types labeled by $\mathbf{0}, \mathbf{\frac12}, \mathbf{1}, \ldots \mathbf{\frac{k}{2}}$. We analyzed the exponents for $O^{q}$ with $q=\mathbf{1}$, which is the non-trivial operator allowed on two $\mathbf{\frac12}$ anyons. Chains of $\mathbf{\frac12}$ anyons coupled by $O^{\mathbf{1}}$ were argued in Ref.~\onlinecite{parameswaranEigenstatePhaseTransitions2017} to approach the Heisenberg spin chain of spins-$\frac12$. We find that $\alpha, \beta, \gamma$ increase monotonically with $k$, approaching limiting values as $k \to \infty$. The limiting value for $\alpha$ is found to be $\log_2 3$, which indeed matches the corresponding result for \sutwo tree states found by Refs.~\onlinecite{protopopovEffectSUSymmetry2017,protopopovNonAbelianSymmetriesDisorder2019}, as $\alpha = \log_2 3$ means that each tree state is connected via $S_i \cdot S_{i+1}$ to $3^{D_{i, i+1}}$ other tree states. However, our limiting values for $\beta, \gamma$ do not match those computed in Ref.~\onlinecite{protopopovEffectSUSymmetry2017} for \sutwo, again suggesting that the thermodynamic limit and the $k \to \infty$ limit do not commute. 

\subsection{Why is $\alpha_q > \beta_q$?}
\label{sec:constraintsonexponents}

In the examples considered above, it seems that we always find the number of matrix elements is growing faster that the size of the matrix elements shrinks. It is natural to ask whether there holds generally. In fact, we note that there is an obvious constraint on how small the matrix elements can be on average, as the total operator norm of $V$ is independent of the basis:
$$
\frac{\Tr (V V^{\dagger})}{\Tr I} = \sum_{ab} \left| V_{ab} \right|^2 / \sum_a 1 = {\rm const.}
$$
Thus the average norm of one column of $V$ is constant. Such a column has $2^{\alpha_q D}$ non-zero matrix elements, so the root mean square average of these is $2^{-\alpha_q D/2}$.
Thus the size of the matrix elements measured by root mean square rather than the mean decays with an exponent $\beta^{(2)}_q = \alpha_q /2$.
This can be confirmed by extending our random matrix calculation to compute the second moment of the matrix element distribution, replacing the $F$-symbols in Eq.~\ref{eq:matrixproductF} with squared $F$-symbols. We did this calculation and found that the root mean square size of the matrix elements scales exponentially in $D$ with exactly this exponent.
This argument gives the constraint that $\beta_q \geq \beta^{(2)}_q = \alpha_q /2$.
While this does not lead to an upper bound on $\beta_q$, Table~\ref{table:groupanalytic} shows that $\beta_q \approx \alpha_q / 2$ in all of the cases computed. As the matrix element distributions, as shown in Fig.~\ref{fig:dihedral}, are somewhat featureless with a shape that does not change drastically with $D$, it seems appropriate that the mean and root mean square average of the matrix elements scale similarly with $D$.

The result seems to suggest that \emph{any} perturbation of fixed magnitude that couples many states --- particularly a growing number of states as system size increases --- necessarily is destabilizing. This is not the case. The local perturbations discussed here are mostly off-diagonal; an operator with larger diagonal contributions to its norm could have smaller off-diagonal matrix elements. More importantly, perturbations of constant size can avoid destabilizing states if there are correlations between the energy denominators and the matrix elements. Only the matrix elements between tree states with nearby energies matter for the purposes of creating resonances. Our expectation is that such matrix elements behave typically as if drawn from these featureless distributions. We examine this point further in the next section.
 
\section{Direct counting of resonances}
\label{sec:directcomputation}

\begin{figure}
\includegraphics[]{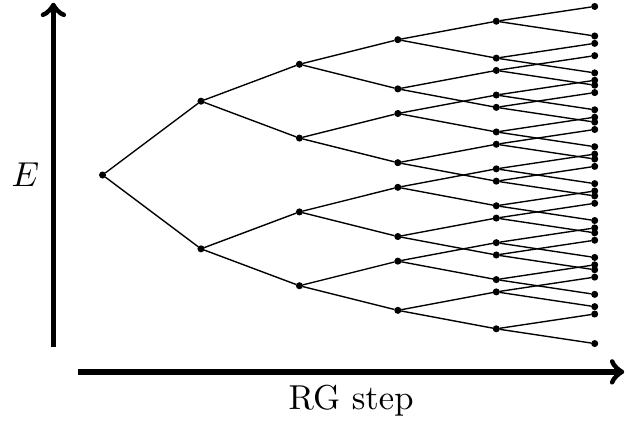}
\caption{Spectral tree generated by RSRG-X}
\label{fig:spectral_tree}
\end{figure}

To show that the off-diagonal matrix elements of local operators indeed cause resonances, we need to consider the energy denominators $E_a - E_b$ of connected states. The hypothesis of Sec.~\ref{sec:resonant_processes} is that small energy denominators occur essentially randomly due to near collisions of the energy of states that differ in many of the locally accessible IOMs.
Some motivation for this hypothesis can be taken from considering the full set of RSRG-X states. The energy scaling for states that only differ by a single IOM takes a stretched exponential form:
$$
\Delta E \sim e^{-\left(L/L_\star \right)^\psi},
$$
with $\psi < 1$, whereas the full set of states has a level spacing that decays exponentially.
Thus, in the full spectrum of states, neighboring states typically differ in many IOMs in the thermodynamic limit where $L \gg L_\star$, as states that differ in just a few IOMs are much further spread in energy. This scenario is depicted in Fig.~\ref{fig:spectral_tree}. The set of locally accessible states are a tiny subset of the set of all states corresponding to picking the same branch of the RSRG process at all but $D$ branchings. Nonetheless, we hypothesize that for $D > D_\star$ those branches will also cross in energy and locally accessible states with nearby energies will differ in many of those $D$ IOMs. Showing this is beyond our analytic arguments. More generally, there may be correlations between the matrix elements and the energy denominators that spoil the resonance counting in Eq.~\ref{eq:NR}.

\begin{figure*}
    \centering
    \includegraphics[width=0.5\linewidth]{{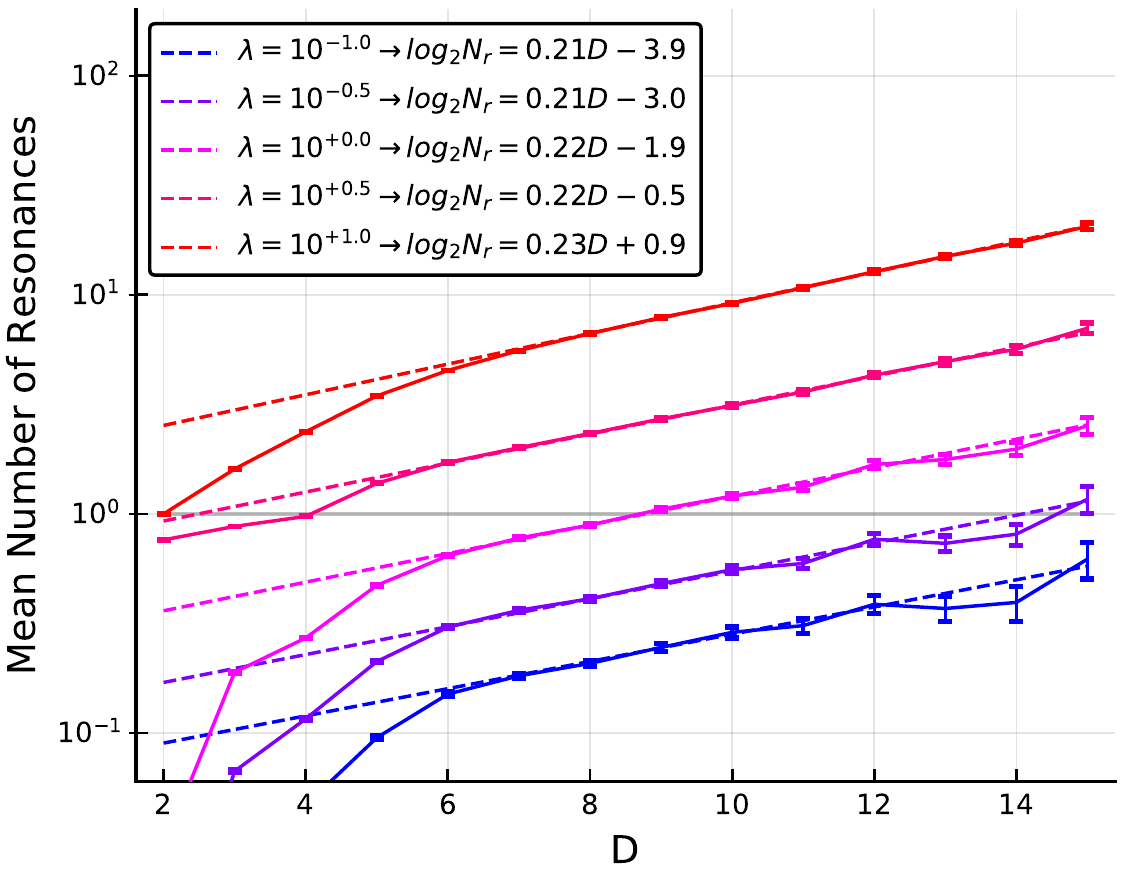}}
    \includegraphics[width=0.3\linewidth]{{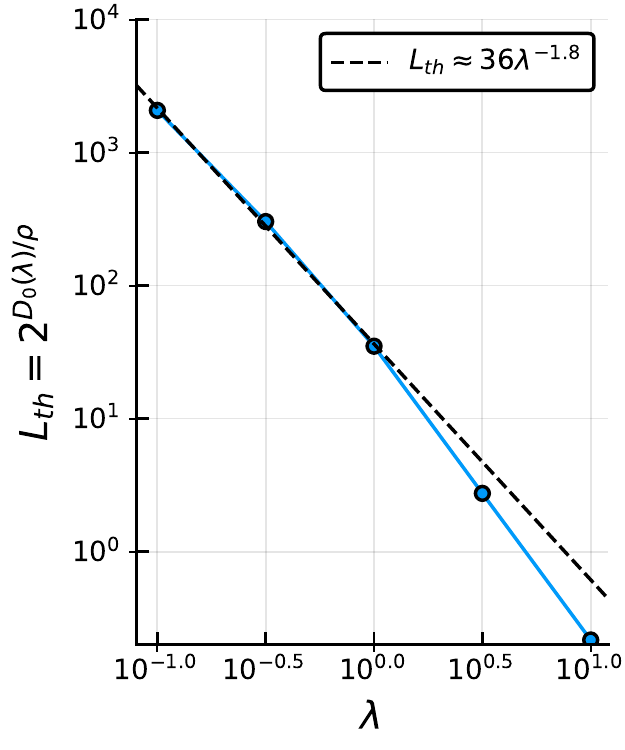}}
    \caption{Direct counting of resonances for $\text{Diag}(H)+ \lambda V$ in the Fibonacci chain where $\text{Diag}(H)$ is obtained using RSRG-X, as a function of $\lambda$ and the local tree depth $D$ at the location of the perturbation. {\it Left}: The $D$ scaling is close to the predictions of Sec.~\ref{sec:sizeandnumber} with the exponent $\gamma_{\tau} = 0.2$ taken from Table~\ref{table:groupanalytic}.  {\it Right: } Thermalization length scale {\it vs} perturbation strength $\lambda$. 
    }
    \label{fig:resonantcounting}
\end{figure*}

We can resolve these questions definitively by numerically computing the energies of locally connected tree states and directly counting the resonances. We do this for the simplest possible QCG phase, that in the disordered Fibonacci anyon chain~\cite{PhysRevLett.98.160409}. The Hamiltonian is 
\begin{equation}
H = \sum_{i} J_i O^{\tau}_{i, i+1},
\end{equation}
with the coefficients $J_i$ randomly sampled from the uniform distribution
$J_i \in [-1, 1]$. The RSRG-X step for this system takes the strongest bond in the system $J_i$ and replaces the corresponding $\tau$ anyons on sites $i, i+1$ with either a singlet or a single $\tau$ anyon. In the former case, the singlet drops out and the anyons on sites $i-1$ and $i+2$ interact via an effective second-order coupling $J_{\text{eff}} = \frac{2}{\varphi^2} \frac{J_{i-1} J_{i+1}}{J_i^2}$. In the later case, the new anyon interacts with its neighbors on either side with first order couplings $J_{\text{eff}} = - J_{i \pm 1}/\varphi$~\cite{bonesteelInfiniteRandomnessFixedPoints2007,fidkowskiPermutationsymmetricCriticalPhases2009,fidkowskiInfiniteRandomnessPhases2008,kangUniversalCrossoverGroundstate2017}.


We generated random disorder instances of such chains with sizes ranging from $L=2^5$ to $L=2^{12}$ anyons. For each disorder instance, we randomly sampled an infinite-temperature random RSRG-X state $\ket{\Psi_a}$. We then computed for each tree state $\ket{\Psi_b}$ connected by a local perturbation $V=O^{\tau}$ the quantitity
$$
\mathcal{G}^V_{ab} = \log \left | \frac{\braket{\Psi_b | V | \Psi_a}}{\braket{\Psi_b | H | \Psi_b} - \braket{\Psi_a | H | \Psi_a}} \right |.
$$
We counted the resonances in the Hamiltonian $H' = \text{Diag}(H) + \lambda V$ by counting the number of connected states with $\mathcal{G}^V_{ab} > - \log \lambda$, as discussed in Sec.~\ref{sec:criteria}.

The result is shown in Fig.~\ref{fig:resonantcounting}. As predicted, the number of such resonances increases exponentially with $D$ with an exponent quite close to the value of $\gamma_\tau \approx 0.2$ computed in the Sec.~\ref{sec:sizeandnumber}. This occurs for all values of $\lambda$ with which we could find enough resonances to get reliable statistical estimates. The regime of exponential growth starts around $D_\star \approx 6$.
In Sec.~\ref{sec:resonant_processes}, our hypothesis of uniformly random level spacing predicted $N_r \propto \lambda$, but the counting here is consistent with a more general form
\begin{equation}
N_r = \lambda^{\zeta} 2^{\gamma_{\tau} D}.
\end{equation}
Most importantly, for any $\lambda$ the trends indicate that there is a corresponding $D$ beyond which resonances proliferate.

We can estimate the associated length scale for thermalization using the result of this resonance counting as input. First we estimate a $\lambda$-dependent $D_0$ at which these resonances proliferate in $\text{Diag}(H)+ \lambda V$ by setting a cutoff in the number of resonances per state necessary to thermalize the system. We arbitrarily set this cutoff at $1$. By using the fits in Fig.~\ref{fig:resonantcounting}, we estimate the threshold $D_0(\lambda)$ and the corresponding length scale $L_{\text{th}}(\lambda) \sim 2^{D_0(\lambda) / \rho}$.
We use the estimate for $\rho$ from Sec.~\ref{sec:resonant_processes}. The result is shown in Fig.~\ref{fig:resonantcounting}.

Finally, to convert this to a length scale for the proliferation of resonances in the unperturbed Hamiltonian, we need to estimate the size of $\lambda$ for the coefficient of $H - \text{Diag}(H)$ along the direction of the most destabilizing operator. As $\lambda$ approaches $0$ with increasing disorder strength, the corresponding thermalization length can be made arbitrarily large.





\section{Conclusion}
\label{sec:conclusion}
We considered the stability of QCG phases in spin chains with non-Abelian symmetry and non-Abelian anyon chains. We mapped the scaling of the size and number of matrix elements of local perturbations to a random matrix product problem. We evaluated the associated Lyapunov exponents for the dihedral groups \Dk{k}, and the \sutwok{k} and \sothreek{k} anyon systems. In all of these cases, the scaling suggests that local perturbations drive resonances that flip many integrals of motion, and that the density of these resonances increases with system size. 

A core distinction for QCG phases from MBL phases is that the number of IOMs that can be flipped with one application of a local operator scales with the tree depth $D_{\text{max}}$ of the RSRG-X tree, with $D_{\text{max}} \sim \log L$. The number of connected states to a given tree state for any local operator is found to scale exponentially in the number of accessible IOMs $D$, while the size of these matrix elements decays exponentially in $D$. The competition between these exponentials always favors the number of connected states, so that for large $D$ the level spacings inevitably become too small to avoid resonant mixing by the matrix elements. The main technical advance of this paper is a method for computing the asymptotic exponential growth of the number and size of the matrix elements for a basis of nearest neighbor operators, which is accomplished by exploiting a nice structure which occurs for these particular matrix elements.

Taken together, the result of this work and Refs.~\onlinecite{protopopovEffectSUSymmetry2017,protopopovNonAbelianSymmetriesDisorder2019} strongly suggests that QCG phases are unstable for every non-Abelian group, discrete or continuous. Systems with discrete non-Abelian symmetries thus must either thermalize or exhibit MBL combined with spontaneous symmetry breaking to an Abelian subgroup.
Additionally, we conclude that QCG phases are unstable in chains of non-Abelian anyons, excluding chains of Majorana anyons or parafermions for which our approach is not predictive. 
Spontaneous symmetry breaking is not a viable option for either these non-Abelian anyon chains or for spin chains with continuous non-Abelian symmetry --- thus, thermalization remains the only possibility in these cases~\cite{potterSymmetryConstraintsManybody2016}.
In all of these systems, the integrals of motion produced by RSRG-X are \emph{approximately} conserved and continue to control the dynamics up to parametrically long time scales $ t \ll t_{\rm th}$, with $t_{\rm th}$ given by Eq.~\eqref{eqtth}. 
Our computation allows us to estimate the time and length scales in which many-body resonances proliferate. We find that these scales are parametrically long as disorder strength increases, and thus the QCG regime continues to be a valid description for the dynamics at strong disorder at practically accessible scales. 
However, our perturbative mechanism for instability may be accompanied by other mechanisms, in which case the QCG description may break down at even smaller scales than we have reported.

While QCGs remain good examples of ``almost'' non-ergodic states up to stretched-exponentially long time scales, our work suggests that they eventually thermalize. Finding genuine (and other long-lived) examples of non-ergodic phases beyond MBL remains a major challenge in the field, and proving the stability of such tentative non-ergodic states might prove even more challenging. Transitions between distinct MBL phases in the case of Abelian symmetries (say in the random transverse-field Ising chain) provide an example non-ergodic state that goes beyond MBL, which is not ruled out by our analysis. However, recent numerical studies suggest that such MBL-MBL transitions are {\em perturbatively} unstable against thermalization~\cite{2020arXiv200809113M, 2020arXiv200808585S}. While the symmetry structure in that case is not enough to explain this instability, the approach considered in this paper (and Refs.~\onlinecite{protopopovEffectSUSymmetry2017,protopopovNonAbelianSymmetriesDisorder2019}) might still provide a useful tool to analyze the stability of the non-interacting Ising transition against adding interactions.


\emph{Acknowledgments}.---The authors thank Wen Wei Ho, Vedika Khemani, Sid Parameswaran and Andrew Potter for useful discussions. This work was supported by the US Department of Energy, Office of Science, Basic Energy Sciences, under Early Career Award No. DE-SC0019168 (R.V.), the Alfred P. Sloan Foundation through a Sloan Research Fellowship (R.V.), and the Swiss National Science Foundation (D.A.).


\bibliography{NonAbelianNonErgodic}

\appendix

\section{Properties of Clebsch-Gordan tensors and tree states}
\label{sec:treestates}
 Tree states for non-abelian spin chains are formulated in terms of the following representation theoretic data:
\begin{itemize}
    \item the set \iA\ of irreducible representations (irreps) of $G$ = $\{ {\bf a}, {\bf b}, {\bf c}, \dots \}$
    \item the identity irrep, which we will label ${\bf 1} \in \iA$
    \item the dimensions of the irreps $d_a$ for each ${\bf a} \in \iA$
    \item the conjugate irrep $\bar{{\bf a}} \in \iA$ for each ${\bf a} \in \iA$
    \item irrep fusion ${\bf a} \times {\bf b} = \sum_c N^{c}_{ab} {\bf c}$ for each ${\bf a}, {\bf b} \in \iA$
    \item Clebsch-Gordan tensors $C(a, b, c, \mu)$ for each ${\bf a}, {\bf b}, {\bf c} \in \iA$, $\mu \in [1, 2, \ldots N^c_{ab}].$
\end{itemize}
The dimension of the representation ${\bf a} \times {\bf b}$ can be computed before and after decomposing into irreps, resulting in the relation
\begin{equation}
d_a d_b = \sum_c N^c_{ab} d_c.
\label{eq:dimensions}
\end{equation}
The Clebsch-Gordan tensors $C(a, b, c, \mu)$ specify the irreducible multiplets of states that mix under the action of $G$ in the tensor product of two irreducible representations.
As all cases we consider have all $N^c_{ab} \in \{0, 1\}$, we omit the fusion multiplicity indices $\mu$ from the rest of the discussion.
For fixed ${\bf a}, {\bf b}$, the $C(a, b, c)$ for ${\bf c} \in {\bf a} \times {\bf b}$ form a complete and orthonormal basis for the states in the Hilbert space $\mathcal{H}_a \otimes \mathcal{H}_b$, which leads to the following relations:
\begin{center}
\begin{tikzpicture}[on grid,
        edge/.style={draw,postaction={on each segment={early arrow=arrowstyle}}},
        lateredge/.style={draw,postaction={on each segment={later arrow=arrowstyle}}}]
    \def \pd{0.6}
    \def \nd{0.5}
    \def \td{0.1}
    \node (center) {};
    \node[left = 2.5*\nd of center] (sigma) {$\nsum[2]\limits_c$};
    \node[above = \nd of center] (a) [circle, draw] {};
    \node[above left = \pd of a] (cl) {};
    \node[above right = \pd of a] (cr) {};
    \path[edge] (cl.north)--(a) node [midway, left=\td] {${\bf a}$};
    \path[edge] (cr.north)--(a) node [midway, right=\td] {${\bf b}$};
    \node[below = of a, circle, draw] (b) {};
    \node[below left = \pd of b] (dl) {};
    \node[below right = \pd of b] (dr) {};
    \path[edge] (b)--(dl.south) node [midway, left=\td] {${\bf a}$};
    \path[edge] (b)--(dr.south) node [midway, right=\td] {${\bf b}$};
    \path[lateredge] (a) -- node [midway, right=\td] {${\bf c}$} (b);
    \node[right = of center] (eq) {$=$};
    
    \node[right = 2.5*\nd of eq] (center2) {};
    \node[above = \nd of center2] (a2) {};
    \node[below = \nd of center2] (b2) {};
    \node[above left = \pd of a2] (cl2) {};
    \node[above right = \pd of a2] (cr2) {};
    \node[below left = \pd of b2] (dl2) {};
    \node[below right = \pd of b2] (dr2) {};
    \path[lateredge] (cl2.north) -- node [midway, left=\td] {${\bf a}$} (dl2.south);
    \path[lateredge] (cr2.north) -- node [midway, right=\td] {${\bf b}$} (dr2.south);
\end{tikzpicture}\,
\begin{tikzpicture}[on grid,
        edge/.style={draw,postaction={on each segment={early arrow=arrowstyle}}},
        lateredge/.style={draw,postaction={on each segment={later arrow=arrowstyle}}}]
    \def \pd{0.6}
    \def \nd{0.5}
    \def \td{0.1}
    \node (center) {};
    \node[above = \nd of center] (a) [circle, draw] {};
    \node[above = \pd of a] (c) {};
    \path[edge] (c.north)--(a) node [midway, right=\td] {${\bf c}$};
    \node[below = of a, circle, draw] (b) {};
    \node[below = \pd of b] (d) {};
    \path[edge] (b)--(d.south) node [midway, right=\td] {${\bf c'}$};
    \path[lateredge] (a) to[out=-30, in=30] node [midway, right=\td] {${\bf b}$} (b);
    \path[lateredge] (a) to[out=-150, in=150] node [midway, left=\td] {${\bf a}$} (b);
    \node[right = of center] (eq) {$=$};
    \node[right = \nd of eq] (center2) {};
    \node[right = \nd of center2] (delta) {$\delta_{c,c'}$};
    \node[above = \nd + \pd of center2] (c2) {};
    \node[below = \nd + \pd of center2] (d2) {};
    \node[below = \nd of center2] (b2) {};
    \path[] (center2) -- node [midway, right=\td] {${\bf c}$} (d2.south);
    \path[lateredge] (c2.north) -- (d2.south);
\end{tikzpicture}
\end{center}
where the Clebch-Gordan tensor $C(a, b, c)$ is represented graphically by 
\begin{center}
\begin{tikzpicture}[on grid,
        edge/.style={draw,postaction={on each segment={early arrow=arrowstyle}}},
        lateredge/.style={draw,postaction={on each segment={later arrow=arrowstyle}}}]
    \def \pd{0.6}
    \def \nd{0.5}
    \def \td{0.1}
    \def \qd{0.6}
    \node (center) {};
    \node[above = \nd of center] (a) [circle, draw] {};
    \node[above = \pd of a] (c) {};
    \path[edge] (c.north)--(a) node [midway, right=\td] {${\bf c}$};
    \node[below = \qd of a] (b) {};
    \node[left = \qd of b] (bl){};
    \node[right = \qd of b] (br){};
    \path[edge] (a) to node [midway, right=\td] {${\bf b}$} (br);
    \path[edge] (a) to node [midway, left=\td] {${\bf a}$} (bl);
\end{tikzpicture}.
\end{center}


For a spin chain with $L$ sites where each site carries an action of the symmetry group $G$ with representation ${\bf r}$, a basis of states can be recursively built for any binary tree shape using Clebsch-Gordan tensors.
The states in the basis are identified by global quantum numbers -- an irrep ${\bf a} \in \iA$ that specifies what type of multiplet the state belongs to and an integer $m \in [1, \ldots, d_a]$ that specifies which state in the multiplet it is -- and additionally by irrep labels ${\bf a_i} \in \iA$ assigned to each leg $i$ of the tree.
The legs at the bottom of the tree are labeled with the irrep ${\bf r}$ --- or, if ${\bf r}$ is reducible, an irrep in the decomposition of ${\bf r}$. Each internal leg must be assigned an irrep compatible with the fusion of the irreps immediately below it.
As an example, the tree state
\begin{center}
  \begin{tikzpicture}[edge/.style={draw,postaction={on each segment={early arrow=arrowstyle}}}]
    \node (TT) at (0, 0.3) {};
    \node[circle, draw] (T) at (0, 0.8) {};
    \node[circle, draw] (A) at (-0.75, 1.75){};
    \node[circle, draw] (B) at (0.375, 1.5) {};
    \node(AA) at (-1., 2.5) {};
    \node(AB) at (-0.5, 2.5) {};
    \node(BA) at (0, 2.5) {};
    \node[circle, draw] (BB) at (0.75, 2) {};
    \node(BBA) at (0.5, 2.5) {};
    \node(BBB) at (1, 2.5) {};

    \path[edge] (T) -- node [midway, right=0.01cm] {${\bf a}, {\bf m}$} (TT.south);
    \path[edge] (A) -- node [midway, below left=-0.05cm] {${\bf a_1}$}(T);
    \path[edge] (B) -- node [midway, below right=-0.05cm] {${\bf a_2}$}(T);
    \path[edge] (BB) -- node [midway, below right=-0.05cm] {${\bf a_3}$}(B);
    \path[edge] (AA.north) -- node [pos = 0.25, left=0.0cm] {${\bf r}$} (A);
    \path[edge] (AB.north) -- node [pos = 0.25, right=-0.04cm] {${\bf r}$} (A);
    \path[edge] (BA.north) -- node [pos = 0.185, left=-0.04cm] {${\bf r}$} (B);
    \path[edge] (BBA.north) -- node [pos = 0.38, left=0.01cm] {${\bf r}$} (BB);
    \path[edge] (BBB.north) -- node [pos = 0.38, right=0.01cm] {${\bf r}$} (BB);
  \end{tikzpicture}
\end{center}
represents a state of $5$ spins transforming as the irrep ${\bf r}$ that globally belongs to an irrep that transforms as ${\bf a}$. As $C(a, b, c)$ only exists if the irrep ${\bf c}$ is part of the decomposition ${\bf a} \times {\bf b}$, which occurs if $N^c_{ab}>0$, this tree represents a state only if 
$$N^{a}_{a_1 a_2} N^{a_1}_{rr}N^{a_2}_{ra_3} N^{a_3}_{rr}>0.$$

As each tree shape generates a complete basis, the states of tree bases with different shapes can be related to each other with a change of basis transformation. For three sites, there are two possible trees. The change of basis matrix between the corresponding two bases is called the $F$-symbol, which is expressed in terms of the Clebsch-Gordan tensors as
\begin{center}
  \begin{tikzpicture}[edge/.style={draw,postaction={on each segment={early arrow=arrowstyle}}},
  lateredge/.style={draw,postaction={on each segment={later arrow=arrowstyle}}}]
    \def \hd{1}
    \def \pd{0.2}
    \def \lrd{0.}
    \def \ad{0.5}
    \def \bd{0.5}
    \def \td{0.1}

    \node (center) {};
    \node (X) at (-\hd, 0) {};
    \node (Y) at (\hd, 0) {};
    \node[circle, draw] (T) at (-\lrd, \hd){};
    \node [above=\pd of T] (TT) {};
    \path[edge] (TT.north) -- node [midway, left=\td] {$d$}(T);

    \node[circle, draw] (B) at (\lrd, -\hd){};
    \node [below=\pd of B] (BB) {};
    \path[edge] (B) -- node [midway, right=\td] {${\bf d}$}(BB.south);

    \node[circle, draw] (L) at (-\ad, -\bd) {};
    \node[circle, draw] (R) at (\ad, \bd) {};

    \path[edge] (T) -- node [midway, above right=\td] {${\bf e}$} (R);
    \path[lateredge] (T) -- (X.center) -- node [midway, left=\td] {${\bf a}$} (L);
    \path[edge] (L) -- node [midway, below left=\td] {${\bf f}$} (B);
    \path[edge] (R) -- node [pos=0.75, right] {${\bf b}$} (L);
    \path[lateredge] (R) -- (Y.center) -- node [pos=0.25, right=\td] {${\bf c}$} (B);

    \node[right of = Y] (eq) {$= F^{abc}_{def}$};
    \node[right of = eq] (center2) {};
    \node[above of = center2] (T2) {};
    \node[below of = center2] (B2) {};
    \path[lateredge] (T2.north) -- node [pos=0.75, right] {${\bf d}$} (B2.south);
  \end{tikzpicture}
\end{center}
The change of basis matrices between any two tree bases on a larger number of sites can be decomposed into a product of $F$-symbols by repeatedly applying the relation 
\begin{center}
  \begin{tikzpicture}[edge/.style={draw,postaction={on each segment={early arrow=arrowstyle}}},
  lateredge/.style={draw,postaction={on each segment={later arrow=arrowstyle}}}]
    \def \hd{1}
    \def \pd{0.2}
    \def \lrd{0.}
    \def \ad{0.5}
    \def \bd{0.5}
    \def \td{0.1}

    \node (center) {};
    \node (X) at (-\hd, 0) {};
    \node (Y) at (\hd, 0) {};
    \node[circle, draw] (T) at (-\lrd, -\hd){};
    \node [below=\pd of T] (TT) {};
    \path[edge] (T) -- node [midway, left=\td] {${\bf d}$}(TT.south);

    \node[circle, draw] (R) at (-\ad, -\bd) {};

    \path[edge] (R) -- node [midway, below left = -0.05cm] {${\bf f}$} (T);
    \path[lateredge] (Y.center) -- node [pos=0.2, left] {${\bf c}$} (T);
    \path[edge] (center.center) -- node [pos=0.35, right=0.2mm] {${\bf b}$} (R);
    \path[edge] (X.center) -- node [pos=0.35, left=0.2mm] {${\bf a}$} (R);

    \node(eq) at (1.5*\hd, -1*\pd - \hd/2){$= \nsum[2]\limits_{\bf e} $};
    \node(eqr) at (4.7*\hd, -1*\pd - \hd/2){$F^{abc}_{def}$};
    \node[right = 3.2cm of center] (center2) {};
    \node (X2) at ($ (center2) + (-\hd, 0) $){};
    \node (Y2) at ($ (center2) + (\hd, 0) $){};
    \node[circle, draw] (T2) at ($ (center2) + (-\lrd, -\hd)$){};
    \node [below=\pd of T2] (TT2) {};
    \path[edge] (T2) -- node [midway, left=\td] {${\bf d}$}(TT2.south);
    \node[circle, draw] (R2) at ($(center2) + (\ad, -\bd)$){};
    \path[edge] (R2) -- node [midway, below right = -0.05cm] {${\bf e}$} (T2);
    \path[lateredge] (X2.center) -- node [pos=0.12, left] {${\bf a}$} (T2);
    \path[edge] (center2.center) -- node [pos=0.35, left=0.2mm] {${\bf b}$} (R2);
    \path[edge] (Y2.center) -- node [pos=0.35, right=0.2mm] {${\bf c}$} (R2);
  \end{tikzpicture}.
\end{center}

For the anyonic Hilbert spaces considered in this paper, there are equivalent notions of tree states and changes of basis between them using $F$-symbols. The tree states are not built in this case using Clebsch-Gordan tensors --- instead, the non-product Hilbert space of many anyons is defined in terms of the tree states. All relations that we derive in terms of $F$-symbols only for non-Abelian spin chains also work for anyon chains. For more information on anyonic Hilbert spaces, see Refs.~\onlinecite{RevModPhys.80.1083,101143}.

Symmetric operators acting on a single irrep must be proportional to the identity, a fact that is known as Schur's Lemma. By inserting resolutions of the identity for ${\bf a} \times {\bf b}$, we can see that symmetric operators on two irreps ${\bf a}, {\bf b}$ must take the form 
\begin{center}
\begin{tikzpicture}[on grid,
        edge/.style={draw,postaction={on each segment={early arrow=arrowstyle}}},
        lateredge/.style={draw,postaction={on each segment={later arrow=arrowstyle}}}]
    \def \pd{0.6}
    \def \nd{0.5}
    \def \td{0.1}
    \node (center) {};
    \node[left = 2.5*\nd of center] (sigma) {$\nsum[2]\limits_{\bf c} \alpha_c$};
    \node[above = \nd of center] (a) [circle, draw] {};
    \node[above left = \pd of a] (cl) {};
    \node[above right = \pd of a] (cr) {};
    \path[edge] (cl.north)--(a) node [midway, left=\td] {${\bf a}$};
    \path[edge] (cr.north)--(a) node [midway, right=\td] {${\bf b}$};
    \node[below = of a, circle, draw] (b) {};
    \node[below left = \pd of b] (dl) {};
    \node[below right = \pd of b] (dr) {};
    \path[edge] (b)--(dl.south) node [midway, left=\td] {${\bf a}$};
    \path[edge] (b)--(dr.south) node [midway, right=\td] {${\bf b}$};
    \path[lateredge] (a) -- node [midway, right=\td] {${\bf c}$} (b);
    \node[left = of sigma] (eq) {$=$};
    \node[left = 1.5*\nd of eq] (center2) {$O^{\alpha}_{ab}$};
\end{tikzpicture},
\end{center}
that is they must be linear combinations of projectors $P^c_{ab}$ onto a fixed combined irrep ${\bf c}$. The constants are fixed by projecting: $$ \alpha_c = \frac1{d_c} \Tr (P^c O^\alpha). $$
Using this formula and comparing to the definition of the $F$-symbol above, we see that the basis of operators used in the text are
\begin{center}
\begin{tikzpicture}[on grid,
        edge/.style={draw,postaction={on each segment={early arrow=arrowstyle}}},
        lateredge/.style={draw,postaction={on each segment={later arrow=arrowstyle}}}]
    \def \pdl{1.}
    \def \pds{0.5}
    \def \nd{0.55}
    \def \td{0.1}
    \node (center) {};
    \node[above left = 0.5*\nd and \nd of center] (a) [circle, draw] {};
    \node[above = \pds of a] (cu) {};
    \node[below = \pdl of a] (cd) {};
    \path[edge] (cu.north)--(a) node [midway, left=\td] {${\bf a}$};
    \path[lateredge] (a) -- (cd.south) node [midway, left=\td] {${\bf a}$};
    \node[below right = 0.5*\nd and \nd of center, circle, draw] (b) {};
    \node[above = \pdl of b] (du) {};
    \node[below = \pds of b] (dd) {};
    \path[lateredge] (du.north) -- (b) node [midway, right=\td] {${\bf b}$};
    \path[edge] (b)--(dd.south) node [midway, right=\td] {${\bf b}$};
    \path[edge] (a) -- node [midway, above=\td] {${\bf q}$} (b);
    
    \node[left = of center] (eq) {$=$};
    \node[left = 1.5*\nd of eq] (center2) {$O^q_{ab}$};
    \node[right = 6.6*\nd of eq] (center3) {$=\nsum[2]\limits_{\bf c}  \frac{1}{d_c} F^{aqb}_{cba} P^{c}_{ab}.$};
\end{tikzpicture}
\end{center}

To derive Eq.~\ref{eq:matrixelement} for the matrix elements of the two-site operator $O^q_{ab}$ in a tree basis, one can use a sequence of $F$-moves to change the tree shape into one in which the two spins on which the operator acts fuse immediately.

\section{Exact Lyapunov exponents in select cases}
\label{sec:exact}

The Lyapunov exponents of the random matrix problems described in Section~\ref{sec:randommatrixproducts} can be computed exactly if the matrices involved can be simultaneously diagonalized or if they have a common leading eigenvector.

By \ref{eq:dimensions}, the matrices $n^a$ with matrix elements 
$
(n^a)_{bc} = N^c_{ab}
$
all have a common eigenvector $\vec{d}$ with components $d_a$ the dimensions of the irreps:
$$
(n^a \vec{d})_b = \sum_c N^c_{ab} d_c = d_a d_b = d_a (\vec{d})_b.
$$
By the Perron-Frobenius theorem, this is the unique largest eigenvector of the matrices $n^a$, as $n^a$ has all nonnegative entries~\cite{preskillLectureNotesQuantum2004}.
Repeated multiplications of any starting vector by a sequence of $n^a$  various $a$ leads to convergence to a multiple of $\vec{d}$:
$$
\left( \prod_{i=1}^D n^{a_i} \right) \vec{v} \to \Lambda(\{a_i\}) \vec{d}.
$$
Similarly, the product of matrices itself converges to a multiple of the projector onto $\vec{d}$:
$$
\prod_{i=1}^D n^{a_i}  \to \Lambda(\{a_i\}) \ket{d}\bra{d}.
$$
Upon multiplying one more matrix $n^{a_{D+1}}$, the magnitude $\lambda$ grows by a factor of the leading eigenvalue $d_{a_{D+1}}$ with probability $p_a$. Thus asymptotically the magnitude $\Lambda$ grows as 
$$
\Lambda \sim \prod_{i=1}^D d_{a_i} = \prod_a d_a^{D p_a}.
$$
Defining the Lyapunov exponent as $\lambda = \frac{1}{ D} \log_2 \Lambda $, we see that 
$$\lambda \to \sum_a p_a \log_2 d_a.$$

Similarly, we can compute the Lyapunov exponent for other random products where the matrices have a simultaneous leading eigenvector. For $\Fib$, the probabilities $p_a$ are 
$$ p_1 = \frac{1}{2+\varphi}, \quad p_{\tau} = \frac{1+\varphi}{2+\varphi}$$ and the matrices for the $M^{\mathbf{x}}$ random matrix product are
\begin{equation*}
m^{\mathbf{1}} = \left(\begin{matrix}
1 & 0 & 0\\
0 & 1 & 0\\
0 & 0 & 1
\end{matrix}\right), \quad 
m^{\mathbf{\tau}} = \left(\begin{matrix}
0 & 1 & 1\\
1 & 0 & 1\\
1 & 1 & 1
\end{matrix}\right).
\end{equation*}
The leading eigenvalues of $m^{\mathbf{1}}, m^{\tau}$ are $1, 1 + \sqrt{2}$ and clearly involve a common eigenvector as the matrices commute. This leads to the computed value
$$
\mu = \frac{1+\varphi}{2+\varphi} \log_2 \left( 1 + \sqrt{2} \right).
$$
Similarly, the matrices for the $F^{\mathbf{x}}$ random matrix product are
\begin{equation*}
f^{\mathbf{1}} = \left(\begin{matrix}
1 & 0 & 0\\
0 & 1 & 0\\
0 & 0 & 1
\end{matrix}\right), \quad 
f^{\mathbf{\tau}} = \left(\begin{matrix}
0 & \frac{1}{\varphi} & \frac{1}{\varphi^{\frac12}}\\
1 & 0 & 1\\
1 & \frac{1}{\varphi^{\frac12}} & \frac{1}{\varphi}
\end{matrix}\right).
\end{equation*}
The exact expression for the leading eigenvalue of $f^{\mathbf{\tau}}$ can be computed using Mathematica, giving the following expression for the exponent governing the growth of the random matrix product
$$
\\\xi =
\frac{1+\varphi}{2+\varphi} \log_2 \frac{\varphi^{\frac12} + 1 + \sqrt{1 - 2 \varphi^{\frac12} + \varphi + 8\varphi^{\frac32}}}{2 \varphi}.$$

The same method can also give the exponents for \Dthree. The probabilities are 
$$p_{\mathbf{1}} = p_{\mathbf{-1}} = 1/6, \quad p_{\mathbf{2}} = 2/3.$$
The matrices for the $M^{\mathbf{x}}$ random matrix product are
\begin{equation*}
m^{\mathbf{-1}} = \left(\begin{matrix}
0 & 1 & 0 & 0 & 0\\
1 & 0 & 0 & 0 & 0\\
0 & 0 & 0 & 1 & 0\\
0 & 0 & 1 & 0 & 0\\
0 & 0 & 0 & 0 & 1
\end{matrix}\right), \quad 
m^{\mathbf{2}} = \left(\begin{matrix}
0 & 0 & 1 & 1 & 1\\
0 & 0 & 1 & 1 & 1\\
1 & 1 & 0 & 0 & 1\\
1 & 1 & 0 & 0 & 1\\
1 & 1 & 1 & 1 & 0
\end{matrix}\right).
\end{equation*}
$m^{\mathbf{1}}, f^{\mathbf{1}}$ are identity matrices.
These matrices commute and have a common eigenvector with eigenvalues $1$, $1+\sqrt{5}$ of $m^{\mathbf{-1}}, m^{\mathbf{2}}$ respectively.
Similarly for $F$ the matrices are
\begin{equation*}
f^{\mathbf{-1}} = \left(\begin{matrix}
0 & 1 & 0 & 0 & 0\\
1 & 0 & 0 & 0 & 0\\
0 & 0 & 0 & 1 & 0\\
0 & 0 & 1 & 0 & 0\\
0 & 0 & 0 & 0 & 1
\end{matrix}\right), \quad 
f^{\mathbf{2}} = \left(\begin{matrix}
0 & 0 & \frac12 & \frac12 & \frac1{\sqrt{2}}\\
0 & 0 & \frac12 & \frac12 & \frac1{\sqrt{2}}\\
1 & 1 & 0 & 0 & 1\\
1 & 1 & 0 & 0 & 1\\
1 & 1 & \frac1{\sqrt{2}} & \frac1{\sqrt{2}} & 0
\end{matrix}\right).
\end{equation*}
The common leading eigenvector has eigenvalues of $1, 1 + \sqrt{1 + 4\sqrt{2}}$ for $f^\mathbf{-1}, f^\mathbf{\tau}$ respectively.

These examples are unique in that they have only $1$ non-Abelian irrep or anyon with $d_a > 1.$ In all of the other examples, the $m$ and $f$ matrices do not commute and the Lyapunov exponent must be computed as described in Sec.~\ref{sec:randommatrixproducts}. 
%

%




\end{document}